\documentclass[12pt]{article}
\usepackage{fullpage}

\usepackage{authblk}

\usepackage{amsmath,amssymb,mathtools}
\allowdisplaybreaks[1]  

\usepackage{derivative}

\usepackage[hidelinks,bookmarks=false]{hyperref}
\usepackage{cleveref}

\usepackage[style=nature]{biblatex}
\DeclareFieldFormat{pages}{\mkfirstpage{#1}}
\AtEveryBibitem{\clearfield{month}}
\AtEveryBibitem{\clearfield{day}}

\usepackage{xspace}

\usepackage{graphicx}
\graphicspath{{figures/}}

\usepackage{siunitx}
\usepackage{float}

\usepackage{subcaption}

\title{Perspective: An outlook on fluorescence tracking}

\author[1,2]{Lance W.Q. Xu}
\author[1,2,3,*]{Steve Pressé}

\affil[1]{Center for Biological Physics, Arizona State University, Tempe, AZ, USA}
\affil[2]{Department of Physics, Arizona State University, Tempe, AZ, USA}
\affil[3]{School of Molecular Sciences, Arizona State University, Tempe, AZ, USA}
\affil[*]{Corresponding author (email: spresse@asu.edu)}

\date{}

\begin{document}
\maketitle

\begin{abstract}
  Tracking single fluorescent molecules has offered resolution into dynamic molecular processes at the single-molecule level. This perspective traces the evolution of single-molecule tracking, highlighting key developments across various methodological branches within fluorescence microscopy. We compare the strengths and limitations of each approach, ranging from conventional widefield offline tracking to real-time confocal tracking. In the final section, we explore emerging efforts to advance physics-inspired tracking techniques, a possibility for parallelization and artificial intelligence, and discuss challenges and opportunities they present toward achieving higher spatiotemporal resolution and greater computational and data efficiency in next-generation single-molecule studies.
\end{abstract}

\section{Introduction}
\label{sec:intro}
Dynamics has been a central notion to the physical sciences since at least the time of Aristotle\autocite{ROVELLI2015Aristotle}. As our understanding of the natural world has progressed, so has our ability to derive insight from trajectories of physical objects starting from the deterministic evolution of macroscopic celestial bodies---stars and planets---enabling Kepler to formulate his laws of planetary motion\autocite{Kepler2015Astronomia}.

Tracking has since been applied to nearly every physical scale: from celestial bodies to everyday macroscopic objects such as vehicles and humans\autocite{Yan2022Towards, Kadam2024Object}, to smaller tracer particles in fluid dynamics\autocite{Adrian2011Particle, Dabiri2019Particle, Raffel2018Particle, Barta2024proPTV}, cells\autocite{Maska2023Cell, Emami2021Computerized, Ershov2022TrackMate}, biomolecules including proteins\autocite{Wirth2023MINFLUX} and nucleic acids\autocite{Pitchiaya2019Dynamic, Xia2025single}, and even charged particles in high-energy physics experiments\autocite{Ai2022Common, DiMeglio2024Quantum}. Given the broad variability in experimental setups and distinct physical principles governing each regime, the breadth of tracking methodologies reflects each scale's unique context.
Yet, despite the diversity of systems and techniques, most tracking methodologies share common goals: precisely determining locations of tracked objects with as high a spatial resolution as allowed and capturing these locations at frequent intervals reflecting the temporal resolution of the detection apparatus.

In this perspective, we focus on fluorescent single-particle tracking (SPT). This powerful tool has advanced numerous areas of biology, from the dynamics within biomolecular condensates\autocite{Abyzov2022Conformational, Galvanetto2023extreme} and immune receptor dynamics\autocite{Bi2021zar1} to mRNA transport and localization\autocite{Pitchiaya2019Dynamic, Xia2025single}, toward broader cellular processes\autocite{Simon2024guide}.

Although SPT in biological systems represents just one tracking application, many key challenges SPT faces are encountered across tracking methodologies, in fluorescence, challenges include, but are not otherwise limited to, typically low signal-to-noise ratios\autocite{Balasubramanian2023}, difficulty inherent in simultaneously tracking multiple particles\autocite{chenouard2014objective}, particles exhibiting diverse types of random walks\autocite{lee2017unraveling, MunozGil2021Objective}, out-of-focus motion similar that can be confounded with blurry background as well as both appearance and disappearance of particles due to photophysical events\autocite{lee2017unraveling}, optical aberrations\autocite{Ji2017Adaptive, Park2023Label}, and motion blur induced by fast-diffusing particles over an exposure\autocite{Flors2007Stroboscopic, Elf2007Probing, Heckert2022Recovering}.

In this perspective, while not touching upon all key issues above, we briefly review the historical development of SPT, outline some common principles underlying its methodologies, and examine key techniques toward achieving high spatiotemporal resolution in the face of these challenges. We also highlight persistent limitations and explore the promise of emerging analyses, minimizing approximations and means to avoid computational bottlenecks.

\section{Brief history of SPT}
One of the earliest examples of single-particle tracking (SPT) can be traced to Jean Baptiste Perrin's pioneering work of the early 1900s, in which he studied the motion of colloidal particles to experimentally validate Albert Einstein's theoretical explanation of Brownian motion\autocite{chaudesaigues1908mouvement, perrin2005brownian, Einstein1905Ueber}. Perrin manually tracked the trajectories of microscopic particles in these studies by projecting their images onto a sheet of paper and marking their positions by hand. At this stage, spatial resolution was limited by the accuracy of visual inspection and manual annotation, while temporal resolution was approximately \qty{30}{\second}\autocite{perrin2005brownian, Perrin2022Atoms}. This manual approach remained used for the subsequent decades\autocite{nordlund1914neue, Kappler1931Versuche}. Still, it was fundamentally constrained by the inability of the human eye to resolve submicron-scale particles and by the impracticality of capturing the rapid motion of fast-diffusing species. As a result, manual tracking proved inadequate for studying dynamic biological processes unfolding at shorter timescales\autocite{Midtvedt2021Quantitative}.

More generally, tracking through correlative analysis of tracer particle intensities, with tracer particles often numbering in the thousands, itself led to the development of particle image velocimetry (PIV) in the 1980s, which in turn expanded our ability to visualize and quantitate fluid velocity fields\autocite{Adrian2005Twenty, Adrian2011Particle, Raffel2018Particle}. Unlike SPT, focusing on trajectories of individual particles, PIV monitors the displacement of patterns of particle ensembles to infer flow dynamics\autocite{Scharnowski2020Particle, Raffel2018Particle, Adrian2011Particle}. Nevertheless, many underlying principles in PIV and similar techniques, such as particle tracking velocimetry (PTV), parallel those in SPT, including image-based detection, motion analysis, and trajectory reconstruction.

By the late 80s, electronic advancements enabled particle tracking automation. For instance, Geerts \textit{et al.} developed an early form of SPT and applied it to monitor individual gold nanoparticles on the surface of living cells using differential interference contrast microscopy\autocite{geerts1987nanovid}. Their approach replaced manual tracking with an automated algorithm that identifies particles and links them across frames. 

Specifically, colloidal gold particles were rendered visible by their ability to scatter light outside the optical system's aperture. In the initial frame, pixel clusters corresponding to individual particles were identified based on intensity, and each particle's position was computed using the center of mass of its associated pixel cluster, an approach foreshadowing the principles of later single-molecule localization methods\autocite{Lelek2021single}. 

In subsequent frames, the algorithm predicted each particle's new location based on its previously identified position by searching within a small square region centered at that location. If the detected cluster touched the boundary of the search area, the region was incrementally expanded until all potential clusters were fully enclosed. Among the detected clusters, the one whose center of mass was closest to the prior position was selected as the particle's updated location. This approach also introduced a threshold on the maximum allowable displacement per frame, which has since been widely adopted in modern SPT algorithms\autocite{chenouard2014objective, tinevez2017trackmate}.

A core idea from Geerts \textit{et al.}'s work, that new particle positions can be predicted based on previous ones, served as a basis for later refinements. While early implementations used relatively simple square search regions centered at prior positions, modern methods employ more sophisticated approaches such as Kalman filtering\autocite{Presse2023data}, which integrates noisy measurements over time to provide statistically optimized position estimates\autocite{chenouard2014objective, shen2017single, Hou2020Real}.

Geerts \textit{et al.}'s work also introduced a fundamental two-step framework for widefield SPT: first, each frame in a recorded image stack was analyzed independently to localize particle locations (localization), and second, these locations were linked across frames to reconstruct trajectories (linking). Later, SPT methods incorporated diverse localization and linking strategies to accommodate different experimental conditions, including combinatorially growing linking possibilities over a frame stack\autocite{lee2017unraveling}. This sequential approach and threshold setting remain the foundation of many contemporary SPT algorithms\autocite{jaqaman2008robust, chenouard2013multiple, tinevez2017trackmate, Nguyen2023recent, Roudot2023u}.

For example, in 1996, Crocker and Grier refined Geerts \textit{et al.}'s method by developing a mainly setup-agnostic  algorithm\autocite{CROCKER1996methods} later known as the Crocker-Grier algorithm. This algorithm first identified local brightness maxima within each image as candidate particle locations. These locations were then refined by computing a weighted centroid of the surrounding pixels, adjusting the location for subpixel accuracy. To link these refined locations across frames, the algorithm determined the most likely set of \textit{M} correspondences between \textit{M} particle locations in two consecutive frames, assuming non-interacting Brownian motion. This linking procedure effectively represented a maximum likelihood estimate under Brownian motion between adjacent frames, a foundational step toward more comprehensive, global physics-inspired tracking frameworks.

Thresholding was not initially limited to predicting particle locations. Indeed, intensity thresholding was typically used to enumerate the number of particles (deduce particle count). As such, SPT efforts have largely been inherently parametric. However, as elaborated in later sections, the ability to infer the number of particles, a hallmark of nonparametric methods, becomes increasingly crucial in 3D settings, where particles may evolve in and out of focus and thus ``disappear'' from certain frames.

\section{Fluorescent SPT}
The notion of being limited in spatial resolution by the wavelength of emitted light and the numerical aperture of the collection objective\autocite{Rayleigh01101879, abbe1882relation} was ultimately overcome first by optimizing tools for localization in static samples\autocite{Hell1994breaking, Betzig1995proposed, Dickson1997onoff, Klar2000Fluorescence, Gustafsson2000Surpassing, Heintzmann2002Saturated, Lidke2005Superresolution, Betzig2006imaging, Rust2006subdiffraction, HESS2006ultra, Lemmer2008spdm, Heilemann2008Subdiffraction, Jungmann2010Single, Rego2012Nonlinear, Prakash2022Super} and subsequently generalizing these tools for dynamical samples\autocite{Manley2008High, diezmann2017three, Qiao2022Rationalized, Sgouralis2024bnptrack, Qiao2025neural} yielding, under some conditions, spatial precision of \qtyrange{e1}{e2}{\nano\m}.

Here, we skip over many of the impressive key innovations initiated in static samples and jump straight to instances of widefield SPT.

\subsection{Widefield fluorescent offline SPT}
Widefield fluorescent offline SPT is one of the most common and straightforward approaches to tracking single particles\autocite{jaqaman2008robust, Manley2008High, tinevez2017trackmate, chenouard2013multiple, chenouard2014objective, Zagato2014SPT, lee2017unraveling, Wasim2018SPT, Lionnet2021SMT, Roudot2023u, Nguyen2023recent, Sgouralis2024bnptrack, Simon2024guide} with ``widefield'' indicating that the entire sample is illuminated and imaged. Here ``offline'' refers to SPT performed post-acquisition rather than in real-time. A key advantage of widefield fluorescent offline SPT is its ability to simultaneously track multiple particles\autocite{chenouard2014objective}, particularly useful in studying interactions between particles, such as G-protein-coupled receptors undergoing dynamic equilibrium between monomers and dimers\autocite{kasai2014single}, and RNA puncta fusing into larger granules\autocite{Xia2025single}.

\begin{figure}[ht]
  \centering
  \includegraphics[width=\linewidth]{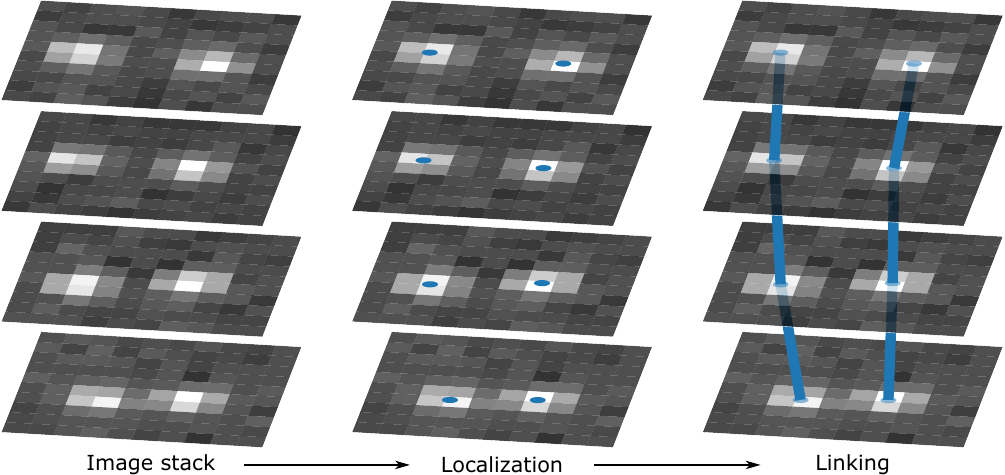}
  \caption{Widefield offline SPT methods process image stacks by first performing a localization step to identify particle positions within each frame. The number of localized particles retained is typically controlled by user-defined thresholds, specific to each method, such as spot quality, radius, or total intensity. These localized positions are subsequently passed to a linking step, which connects them across frames to reconstruct complete particle trajectories\autocite{chenouard2014objective, lee2017unraveling, shen2017single, tinevez2017trackmate}}.
  \label{fig:one}
\end{figure}

Nearly all SPT methods within this category, including those extensively reviewed and compared in Refs.\autocite{chenouard2014objective,lee2017unraveling,shen2017single}, follow the conventional two-step modular process: first, localizing particles within each frame, and second, linking these locations across frames to reconstruct trajectories, as depicted in \cref{fig:one}. Indeed, both of these steps originated from tasks to be solved in static localization problems.

Indeed, many localization techniques initially developed for single-molecule localization microscopy were directly leveraged, lock-stock-and-barrel, to the localization step of SPT frame-to-frame. Concretely, most of these approaches, whether for static or dynamic samples, operate by modeling a single particle's point spread function (PSF) and identifying its center. Common strategies include intensity-weighted centroid calculations\autocite{CROCKER1996methods, Feng2007Accurate, Cheezum2001Quantitative}, the fluoroBancroft algorithm\autocite{Andersson2008Localization}, the radial symmetry method\autocite{Parthasarathy2012Rapid}, and PSF fitting using least-squares or maximum likelihood estimation to achieve sub-pixel precision\autocite{Cheezum2001Quantitative, Thompson2002precise, Yu2011fast, tinevez2017trackmate, Lelek2021single}. More recently, deep learning-based methods have emerged, offering automated localization capabilities and improved robustness under challenging imaging conditions\autocite{Ouyang2018deep, Speiser2021Deep, Faraz2022Deep, Bi2025sptnet}. Each method offers trade-offs between computational efficiency, localization accuracy, and robustness to noise. For instance, PSF fitting provides the highest accuracy, assuming a well-calibrated PSF with minimal variations\autocite{Deng2009Effect, Liu2024Universal}, but can be computationally expensive, while centroid-based methods are faster but less accurate\autocite{shen2017single}.

Similarly, various strategies have been developed for the linking step, also, perhaps counter-intuitively, used for improving the spatial resolution of static samples \autocite{Cox2011Bayesian, Jungmann2016Quantitative, Fazel2019Bayesian} as blinking molecules frame to frame are re-assigned to individual molecules. Common linking approaches include nearest-neighbor linking\autocite{Celler2013single, CROCKER1996methods}, linear assignment problem solvers\autocite{jaqaman2008robust}, multiple hypothesis tracking\autocite{chenouard2013multiple}, and continuous energy minimization\autocite{Milan2014continuous}. More recently, deep-learning-based approaches have been introduced to improve tracking robustness and adaptability to different experimental conditions\autocite{Spilger2021deep, Bi2025sptnet}.

Beyond its obvious advantage in leveraging existing tools, another advantage of the two-step localization and linking process is its speed and adaptability. By decoupling these steps, different tools could be mixed and matched to optimize performance based on experimental conditions or computational constraints. Furthermore, linking algorithms can be integrated with image segmentation techniques to extend tracking capabilities beyond single particles, facilitating applications such as cell tracking\autocite{Ershov2022TrackMate}.

Yet speed and adaptability achieved through modularity present a compromise. Linking relies on localization accuracy, and localization relies on particle number determination (and discrimination from background). While Kalman filtering technology\autocite{chenouard2014objective, shen2017single,lee2017unraveling, Hou2020Real, Presse2023data} may avoid separating localization and linking as independent steps, it cannot in itself learn particle numbers. Moreover, the linearity of the filter is only assured for simple noise (Gaussian) models, rendering it difficult to generalize to realistic cases. As an example, sub-pixel localization necessarily requires an integration of the intensity over a pixel, which immediately introduces complexities to the simple Kalman filter\autocite{lee2017unraveling}. 

While intensity thresholds can be used to approximately enumerate particles\autocite{chenouard2014objective, tinevez2017trackmate}, setting these thresholds manually introduces significant challenges to reliable tracking. Too strict a threshold may introduce selection bias by excluding valid particles, thereby reducing tracking efficiency, while too lax a threshold may result in interpreting noise as a particle\autocite{chenouard2014objective, Sgouralis2024bnptrack}. Fundamentally, this is not only the challenge of SPT, but also a limitation of all techniques based on maximum likelihood\autocite{Liu2025Noise}. This trade-off becomes particularly problematic as the number of tracked particles increases and more threshold parameters are invariably introduced. In such cases, carefully tuning thresholds to balance detection sensitivity and specificity becomes highly time-consuming and may fail to generalize across diverse experimental conditions.

\begin{figure}[ht]
  \centering
  \includegraphics{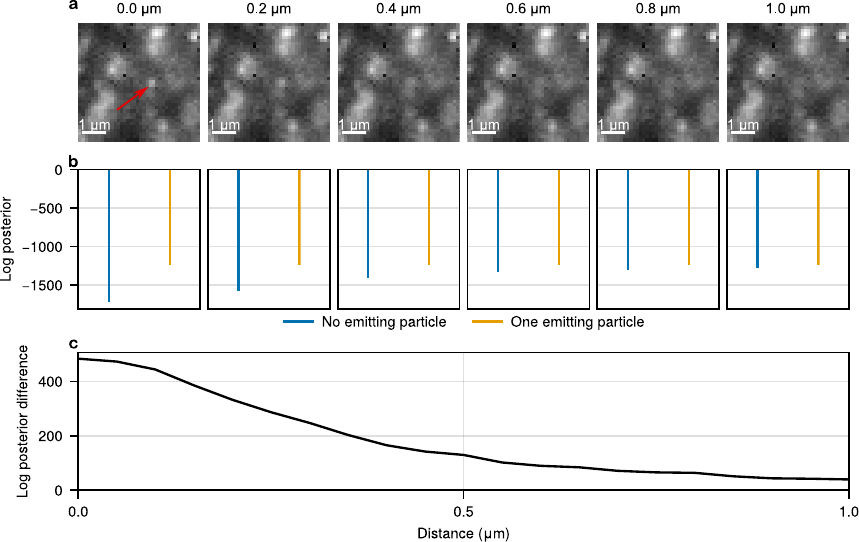}
  \caption{Demonstration of ambiguity between background signal and out-of-focus particle using synthetic data. \textbf{a}, Simulated frames showing a single particle, marked by the red arrow, located at varying axial distances from the focal plane, ranging from \qty{0}{\micro\m} to \qty{1}{\micro\m}. Background noise and camera characteristics are based on real experimental conditions\autocite{Sgouralis2024bnptrack}, assuming an EMCCD camera with gain and offset of 100, a numerical aperture of 1.45, refractive index of 1.515, and emission wavelength of \qty{665}{\nano\m}. \textbf{b}, Comparison of log posterior probabilities under two competing hypotheses: the presence of one emitting particle versus no emitting particles. \textbf{c}, Log posterior difference plotted as a function of the particle's axial distance from the focal plane, illustrating the gradual loss of statistical evidence as the particle becomes increasingly out of focus.}
  \label{fig:OOF}
\end{figure}

What is more, these problems are only further exacerbated in 3D as particles enter or exit the field of view (FOV), introducing indeterminacy between what is background versus what is an out-of-focus particle (see \cref{fig:OOF}) or, even worse, as labels undergo photophysical processes such as blinking or bleaching. While these issues remain open problems, features like ``gap closing'' have been introduced in some tracking algorithms\autocite{tinevez2017trackmate}, albeit introducing additional tunable parameters and user-dependent choices.

While particle number determination, localization, and linking remain challenging in dilute scenarios for slow diffusers, tracking is routinely performed\autocite{Simon2024guide}. Ensuring high localization accuracy and, consequently, reliable tracking, requires sufficient photon counts. According to the Cramér-Rao lower bound\autocite{Kay1993fundamentals, Lelek2021single, fazel2024fluorescence}, the accuracy of localization is fundamentally limited by the number of detected photons, emphasizing the importance of adequate signal levels.

But ramping up photon budget itself can only go so far: the detector's fill factor and its quantum yield are all limited by fundamental physical constraints and experimental conditions\autocite{Grimm2021Caveat, Ashoka2023Brightness}. Chemical constraints are real too: higher laser power elevates the risk of photodamage to the sample and, more broadly, the properties of labels\autocite{Balasubramanian2023}. Alternatively, extending exposure time allows more photons to be captured per frame, increases motion blur\autocite{fernandez-suarez_2008, lee2017unraveling, Martens2022, Clarke2022}---a particularly significant issue when tracking fast-diffusing biomolecules.

However, all of these considerations still ignore key non-idealities. As we approach overlapping PSFs typical of crowded environments\autocite{Holden2011, Huang2011Simultaneous, Zhu2012Faster, Fazel2019Bayesian, Li2019Divide, Nehme2020DeepSTORM3D, Sgouralis2024bnptrack}, aberrations in deeper sample contexts\autocite{Ji2017Adaptive, Park2023Label}, and motion blur indistinguishable from background\autocite{fernandez-suarez_2008, lee2017unraveling,Martens2022, Clarke2022}, all bets are off. These open challenges, appreciated by the broader community and tackled in piecemeal fashion by ourselves\autocite{jazani2022computational, Sgouralis2024bnptrack} and many others\autocite{Holden2011, Huang2011Simultaneous, Zhu2012Faster, Fazel2019Bayesian, Li2019Divide, Nehme2020DeepSTORM3D, Martens2022, Clarke2022}, ultimately motivate deeper introspection into the tracking challenge.

\begin{figure}[ht]
  \centering
  \includegraphics[width=170mm]{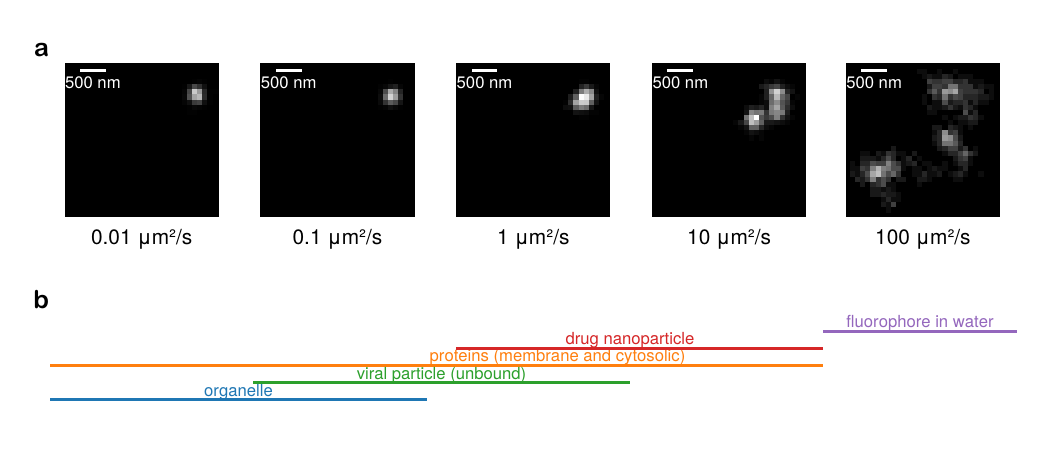}
  \caption{Blurring artefacts in a simulated single molecule labeled with a Cy3B fluorophore (assuming an emission wavelength of \qty{571}{\nano\m}) become increasingly apparent with higher diffusion coefficients under typical camera exposures of \qty{10}{\milli\s}. The pixel size is \qty{100}{\nano\m}, and the objective's numerical aperture is 1.49. Background noise is excluded to emphasize the effect of blurring for illustration. \textbf{a}, Frames show increasing diffusion coefficients from left to right. Contrast is enhanced at higher diffusion coefficients for better visualization. The total number of detected photons per frame remains constant. A single molecule can appear as multiple at higher diffusion coefficients due to blurring artefacts. \textbf{b}, Representative diffusion coefficient ranges for various biological systems\autocite{phillips2012physical}.}
  \label{fig:zero}
\end{figure}

Perhaps the greatest limitation of all presented by modular localization and tracking relies on the notion of localization itself being meaningful. So far, integrative detectors, such as EMCCD and sCMOS cameras used in SPT, have met expectations. Yet this expectation vanishes in the face of single-photon detectors (SPDs). Here, most SPD-based methods either focus on tracking a single particle at a time\autocite{Wirth2023MINFLUX, Chen2023Fluorescence, Scheiderer2025MINFLUX} (discussed further in \cref{sec:RT-SPT}) or rely on summing multiple single-photon frames before applying standard localization techniques similar to those used with integrative detectors\autocite{bucci20244d, Gyongy2018highspeed}. However, this frame-summing approach sacrifices the high temporal resolution that SPDs can deliver. The net result is that it is difficult for offline SPT methods to achieve temporal resolution below milliseconds, limiting their applicability for studying fast-moving particles including, for instance, SARS-CoV-2 virions which iffuse at rates exceeding \qty{5}{\micro\m^2/\s}\autocite{Zbigniew2021SARS}, and cytoplasmic proteins which can reach diffusion coefficients of approximately \qty{10}{\micro\m^2/\s}\autocite{Schavemaker2018}, as illustrated in \cref{fig:zero}.

Additional trade-offs arise in offline SPT when extending tracking to three dimensions, particularly concerning temporal resolution, axial resolution, and axial range. Some 3D tracking methods rely on capturing data across multiple axial planes\autocite{Stelzer2021Light}. While increasing the number of planes can enhance axial resolution and extend the axial range, it inevitably reduces temporal resolution due to the added acquisition time. Alternatively, methods employing engineered PSFs, such as the double-helix PSF\autocite{Pavani2009Three}, can achieve finer axial resolution, but often at the expense of a reduced axial range\autocite{Moerner2015Single, Wang2023Double}.

\subsection{Real-time fluorescence SPT}\label{sec:RT-SPT}
No discussion of SPT would be complete without a note on real-time SPT\autocite{Enderlein2000tracking, Levi20053, Berglund2005Tracking, Katayama2009Real, Hou2019Adaptive, Hou2020Real, Gwosch2020MINFLUX, Zhao2021Leveraging, Schmidt2021MINFLUX,vanHeerden2022real, Johnson2022Capturing, Tan2023Active, Wirth2023MINFLUX, Chen2023Fluorescence, Chen2023Fluorescence,bucci20244d, Scheiderer2025MINFLUX}. Unlike offline fluorescence SPT, which first acquires image stacks and performs tracking post-acquisition, real-time fluorescence SPT actively tracks particles as data are collected. While one may naively think of applying localization algorithms immediately after each image is captured, similar to methods used in offline SPT, this approach is rarely implemented in practice. The primary limitation is the time required to form an image with sufficient photon detections, typically on the order of tens of milliseconds\autocite{vanHeerden2022real}, combined with the computational cost of running standard localization and linking algorithms, often operating on sub-second timescales\autocite{tinevez2017trackmate,chenouard2014objective}. This delay renders conventional localization-based tracking infeasible for freely diffusing particles in real-time.

For the sake of computational efficiency, one strategy toward real-time SPT focuses on computational speedup by utilizing specialized hardware, such as lock-in amplifiers\autocite{Berglund2005Tracking} or field-programmable gate arrays\autocite{Balzarotti2017nanometer, Hou2019Adaptive}, to process data in real-time.

Another common effective strategy for real-time SPT involves restricting the observation volume to a small region, allowing only a single particle to be tracked at a time\autocite{vanHeerden2022real}. Confining the search area eliminates the need for large-scale particle detection across the entire FOV by fundamentally eliminating the computational burden of linking algorithms.

\begin{figure}[ht]
  \centering
  \includegraphics{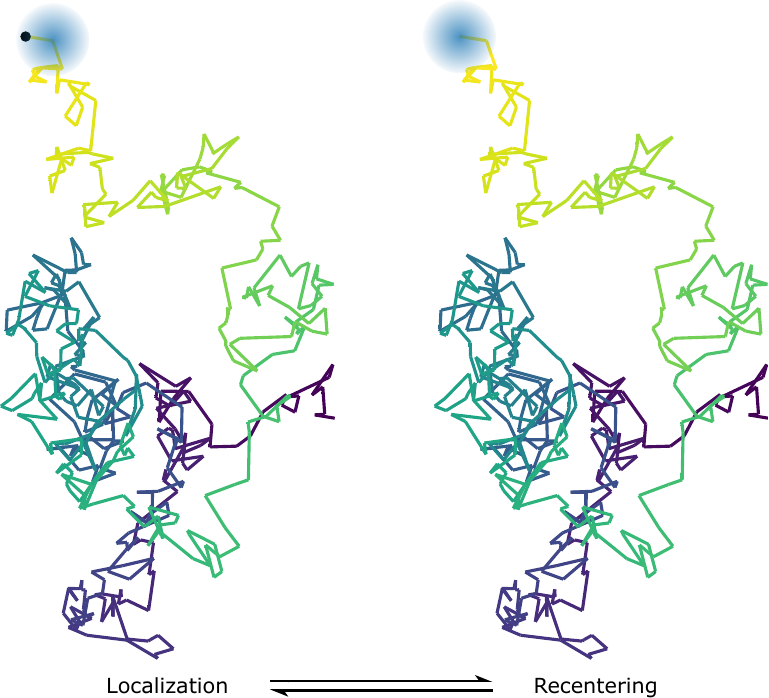}
  \caption{Most real-time SPT methods operate by continuously localizing a single particle in real-time and dynamically adjusting the observation volume to keep the particle centered within the detection volume.}
  \label{fig:two}
\end{figure}

While restricting the observation volume is effective for real-time tracking, it introduces another problem: diffusing particles are prone to escaping the tracking region, limiting the duration over which they can be observed. To address this, many real-time SPT systems employ feedback-driven mechanisms\autocite{Hou2019Adaptive, Hou2020Real, vanHeerden2022real, Tan2023Active, Wirth2023MINFLUX, Chen2023Fluorescence, bucci20244d, Scheiderer2025MINFLUX} dynamically repositioning the observation volume to keep the particle centered, thereby extending the observation window and ensuring more continuous tracking, see \cref{fig:two}.

Moreover, by dynamically repositioning the observation volume, real-time SPT circumvents the trade-offs between temporal resolution, axial resolution, and axial range, which often constrain offline SPT approaches. This capability makes real-time SPT particularly well-suited for tracking particles undergoing three-dimensional diffusion\autocite{Hou2019Adaptive, Hou2020Real, Gwosch2020MINFLUX, Schmidt2021MINFLUX, vanHeerden2022real, Johnson2022Capturing, Tan2023Active, Chen2023Fluorescence, bucci20244d, Scheiderer2025MINFLUX}.

\subsubsection{Confocal real-time fluorescence SPT}
Various techniques have been developed to achieve a small observation volume, many of which rely on confocal microscopy setups in which a focused laser spot rapidly scans the target region. Several scanning schemes have emerged: deterministic scanning\autocite{Enderlein2000tracking, Levi20053, Berglund2006Performance, Hou2020Real}, where the spot follows a fixed, predefined trajectory; constellation scanning\autocite{Perillo2015Deep, Scheiderer2025MINFLUX}, where it dwells at a discrete set of predetermined points; and dynamic scanning\autocite{Andersson2011nonlinear, Ashley2016Tracking}, in which the scan path is adaptively generated in real-time based on particle motion or feedback.

A prominent example of confocal real-time fluorescence SPT is MINFLUX, which combines a donut-shaped excitation PSF with a six-point constellation scanning to achieve spatial resolution below ten nanometers\autocite{Balzarotti2017nanometer, Gwosch2020MINFLUX, Carsten2024MINFLUX, Scheiderer2025MINFLUX}. This exceptional resolution enables the direct visualization of molecular-scale biological processes, such as the unimpeded stepping of kinesin-1 along microtubules\autocite{Wirth2023MINFLUX}.

While reducing the observation volume can enhance localization accuracy\autocite{Vickers2021Information, Gallatin2012Optimal}, expanding the detection volume increases the ability to track fast-diffusing particles by prolonging the time before they escape the observation region\autocite{Berglund2006Performance, Andersson2005Tracking}. This trade-off is crucial in optimizing real-time tracking performance, as improving localization accuracy often comes at the expense of reduced tracking duration for rapidly moving molecules\autocite{McHale2007Quantum, Hou2019Adaptive}.

Additionally, real-time SPT encounters the same fundamental trade-offs as offline SPT, balancing localization accuracy, photodamage, and the maximum diffusion coefficient that can be reliably tracked. The trade-off between localization accuracy and photodamage arises for the same reason as in widefield SPT: higher laser intensities yield more photon detections, improving localization accuracy but increasing the risk of photodamage to biological samples.

\section{Reducing approximations in the physics and dealing with the computational burden of offline SPT}
To help mitigate modeling approximations, we reassess the modular tracking paradigm by re-pitching the SPT problem as one aimed at globally determining the probability distribution over all possible particle tracks given the recorded image frames, denoted as \(\mathbb{P}\left(\mathrm{tracks}\middle|\mathrm{frames}\right)\). Here, ``tracks'' reflects a collection of random variables: the particle count, their spatial locations, and the associations or links between these positions over time. Thus, we can rewrite the original distribution into an equivalent form \(\mathbb{P}\left(\mathrm{tracks}\middle|\mathrm{frames}\right)\equiv \mathbb{P}\left(\mathrm{links},\mathrm{count},\mathrm{locations}\middle|\mathrm{frames}\right)\). This joint probability distribution, which so far is free of modular assumptions, can be further expressed without approximation using conditional probabilities:
\begin{align}
  &\mathbb{P}\left(\mathrm{links},\mathrm{count},\mathrm{locations}\middle|\mathrm{frames}\right)\nonumber\\
  =&\mathbb{P}\left(\mathrm{links}\middle|\mathrm{count},\mathrm{locations},\mathrm{frames}\right)
  \mathbb{P}\left(\mathrm{count}\middle|\mathrm{locations},\mathrm{frames}\right)
  \mathbb{P}\left(\mathrm{locations}\middle|\mathrm{frames}\right).
  \label{eq:spt}
\end{align}

The key distinction between this more global, physics-inspired SPT and conventional offline SPT lies in how they identify the most probable particle tracks from \cref{eq:spt}. The global SPT directly seeks tracks that maximize the full joint probability distribution in \cref{eq:spt}, considering particle count, locations, and links simultaneously. In contrast, traditional offline methods, following the modular localization-and-linking paradigm, decompose this process: they first maximize the probability of particle locations (\(\mathbb{P}\left(\mathrm{locations}\middle|\mathrm{frames}\right)\)) and subsequently estimate particle count and linking based on these localized positions. This stepwise strategy employs a greedy algorithm that optimizes each component in isolation rather than the global objective. As a result, it often fails to identify the globally optimal (most probable) answer and, in some cases, can even yield the worst possible results\autocite{BangJensen2004When}.

This global, Bayesian approach is intrinsically more data efficient than modular approaches; for example, it leverages all frames and all FOVs at each frame at once to identify possible tracks. This approach also immediately remedies an important shortfall, the over-fitting problem, of maximum likelihood by virtue of allowing priors over particle numbers\autocite{Jazani2019Alternative, Tavakoli2020Pitching, jazani2022computational, Sgouralis2024bnptrack}. For example, for confocal applications, Tavakoli \textit{et al.}\autocite{Tavakoli2020Pitching} demonstrated that Bayesian methods can reduce the number of detected photons needed to estimate a molecule's diffusion coefficient by up to three orders of magnitude compared to traditional fluorescence correlation spectroscopy\autocite{Elson1974Fluorescence, Elson2011Fluorescence}. In another study, Jazani \textit{et al.}\autocite{jazani2022computational} introduced an SPT approach that bridges widefield and confocal modalities by utilizing four spatially offset but stationary observation volumes as well as single-photon arrival data. While data-efficient, any method globally using all information to perform tracking will naturally outperform any method using more limited, local information, albeit at computational cost.

Perhaps more importantly, a global Bayesian approach applies not only to integrative detector architectures\autocite{Sgouralis2024bnptrack} but, by virtue of leveraging all spatiotemporal correlations from the data, can be extended to accommodate SPDs, using SPAD array detectors as a representative example\autocite{Xu2025Singlea}. For example, from \cref{eq:spt} above, we can propose particle numbers and associated trajectories consistent with the entirety of the data over the whole field of view over all frames while foregoing the localization paradigm altogether as it is not possible to localize particles, which relies on intensities, from binary SPD output.

\subsection{Theoretical methodology}
As discussed above, achieving global SPT requires considering the full probability distribution defined in \cref{eq:spt}. To facilitate this, we begin by applying Bayes' theorem\autocite{Presse2023data, Bui-Thanh2021Optimality}, yielding:
\begin{equation}
  \mathbb{P}\left(\mathrm{tracks}\middle|\mathrm{frames}\right)=\frac{\mathbb{P}\left(\mathrm{frames}\middle|\mathrm{tracks}\right)\mathbb{P}\left(\mathrm{tracks}\right)}{\mathbb{P}\left(\mathrm{frames}\right)}.
  \label{eq:bayes1}
\end{equation}
In the Bayesian framework, this formulation interprets the target probability distribution \(\mathbb{P}\left(\mathrm{tracks}\middle|\mathrm{frames}\right)\) as the posterior, representing our updated knowledge of particle tracks given the observed frames.  The term \(\mathbb{P}\left(\mathrm{frames}\middle|\mathrm{tracks}\right)\) is the likelihood, encoding the probability of the observed data given a specific set of tracks, while \(\mathbb{P}\left(\mathrm{tracks}\right)\) is the prior, reflecting our assumptions about the tracks before seeing the data. Finally, \(\mathbb{P}\left(\mathrm{frames}\right)\), known as the evidence, serves as a normalization constant

To proceed, we derive each term in \cref{eq:bayes1}, beginning with the likelihood term \(\mathbb{P}\left(\mathrm{frames}\middle|\mathrm{tracks}\right)\). For notational simplicity, we first define the following: \(N\) as the number of frames, \(P\) as the number of pixels per frame, and \(w_n^p\) as the data recorded at pixel \(p\) in frame \(n\). Using this notation, the complete dataset is denoted by \(w_{1:N}^{1:P}=\left\{w_n^p\middle|n=1,\dots,N,p=1,\dots,P\right\}\), and the set of all particle tracks is represented as \(\mathbf{x}_{1:N}^{1:M}\), where \(M\) is the number of particles. The likelihood \(\mathbb{P}\left(\mathrm{frames}\middle|\mathrm{tracks}\right)\) is therefore written as \(\mathbb{P}\left(w_{1:N}^{1:P}\middle|\mathbf{x}_{1:N}^{1:M}\right)\), which, under the assumption that pixel measurements are conditionally independent given the particle positions in each frame, factorizes as:
\begin{equation}
  \mathbb{P}\left(w_{1:N}^{1:P}\middle|\mathbf{x}_{1:N}^{1:M}\right)=\prod_{n=1}^{N}\prod_{p=1}^{P}\mathbb{P}\left(w_n^p\middle|\mathbf{x}_n^{1:M}\right).
\end{equation}

The per-pixel likelihood \(\mathbb{P}\left(w_n^p\middle|\mathbf{x}_n^{1:M}\right)\) depends on four main factors: the optical system's PSF, the pixel area \(A\), the brightness of the particles \(h\), and the number of particles \(M\). The PSF defines how light emitted from a single point source is spatially distributed on the detector surface\autocite{fazel2024fluorescence}, influencing how the signal spreads over neighboring pixels. Particle brightness \(h\) corresponds to the expected photon count at a pixel precisely aligned with the center of a particle's PSF in focus. Given these factors, the total photon contribution from all emitting particles to pixel \(p\) in frame \(n\), denoted as \(u_n^p\), can be calculated as:
\begin{equation}
  u_n^p = h\sum_{m=1}^{M}\iint_{A}\mathrm{PSF}\left(x,y;\mathbf{x}_n^m\right)\odif{x}\odif{y}.
  \label{eq:PSFintegral}
\end{equation}

Given the expected photon count \(u_n^p\), the distribution of the measured signal \(w_n^p\) depends on the type of detector. For integrative detectors, such as sCMOS and EMCCD cameras, \(w_n^p\) is typically modeled using a combination of Gaussian and Gamma distributions to account for photon shot noise and detector noise characteristics\autocite{fazel2024fluorescence, Sgouralis2024bnptrack}. In contrast, SPAD array detectors operate in binary mode, detecting only whether at least one photon has arrived. Consequently, \(w_n^p\) follows a Bernoulli distribution, where the probability of detecting at least one photon is given by \(1 - \mathrm{Poisson}(0; u_n^p)\), with \(\mathrm{Poisson}(0; u_n^p)=e^{-u_n^p}\) representing the probability of zero photon arrivals.

Furthermore, biological specimens frequently introduce sample-induced optical aberrations\autocite{shen2017single, Xu2020Three, fazel2024fluorescence}, causing the PSF in \cref{eq:PSFintegral} to deviate from its idealized theoretical form. While hardware-based approaches such as adaptive optics can correct these aberrations in real time\autocite{Ji2017Adaptive, Rodriguez2018Adaptive}, an alternative strategy is to reconstruct the PSF directly from the imaging data\autocite{Fazel2025Simultaneous}. In this approach, continuous 2D priors are placed on the pupil function's phase and amplitude. This allows the reconstruction to capture fine-scale aberrations beyond those representable by a finite set of Zernike modes, thereby enabling more accurate modeling of complex, sample-specific PSF distortions.

\begin{figure}[h]
  \centering
  \includegraphics{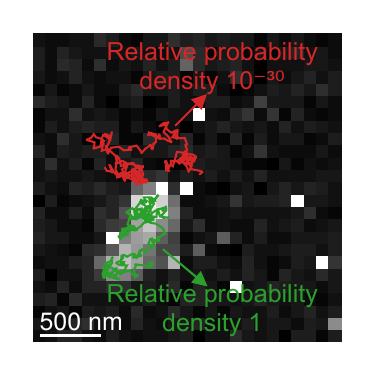}
  \caption{A sampling step of the physics-inspired SPT framework. Particle track proposals closer to the ground truth (red) than those further away (green) are assigned higher posterior probability densities. And these densities ultimately dictate whether the track proposals will be retained. The relative density is defined as the probability density of a given track divided by the probability density of the ground truth track.}
  \label{fig:three}
\end{figure}

Once the posterior distribution \(\mathbb{P}\left(w_{1:N}^{1:P}\middle|\mathbf{x}_{1:N}^{1:M}\right)\) is constructed from the likelihood and prior, a global and physics-inspired SPT framework can proceed by either sampling from this distribution or identifying the most probable set of tracks---known as the \textit{maximum a posteriori} (MAP) estimate or by sampling the posterior. Either task is typically achieved using techniques such as Markov chain Monte Carlo, enabling exploration of the posterior landscape to recover both the tracks and their associated uncertainties\autocite{Presse2023data}. \Cref{fig:three} illustrates a representative step in this sampling process, where two different track proposals are evaluated, and their posterior probabilities compared to determine which, if any, should be retained.

\begin{figure}[ht]
  \centering
  \includegraphics[width=\textwidth]{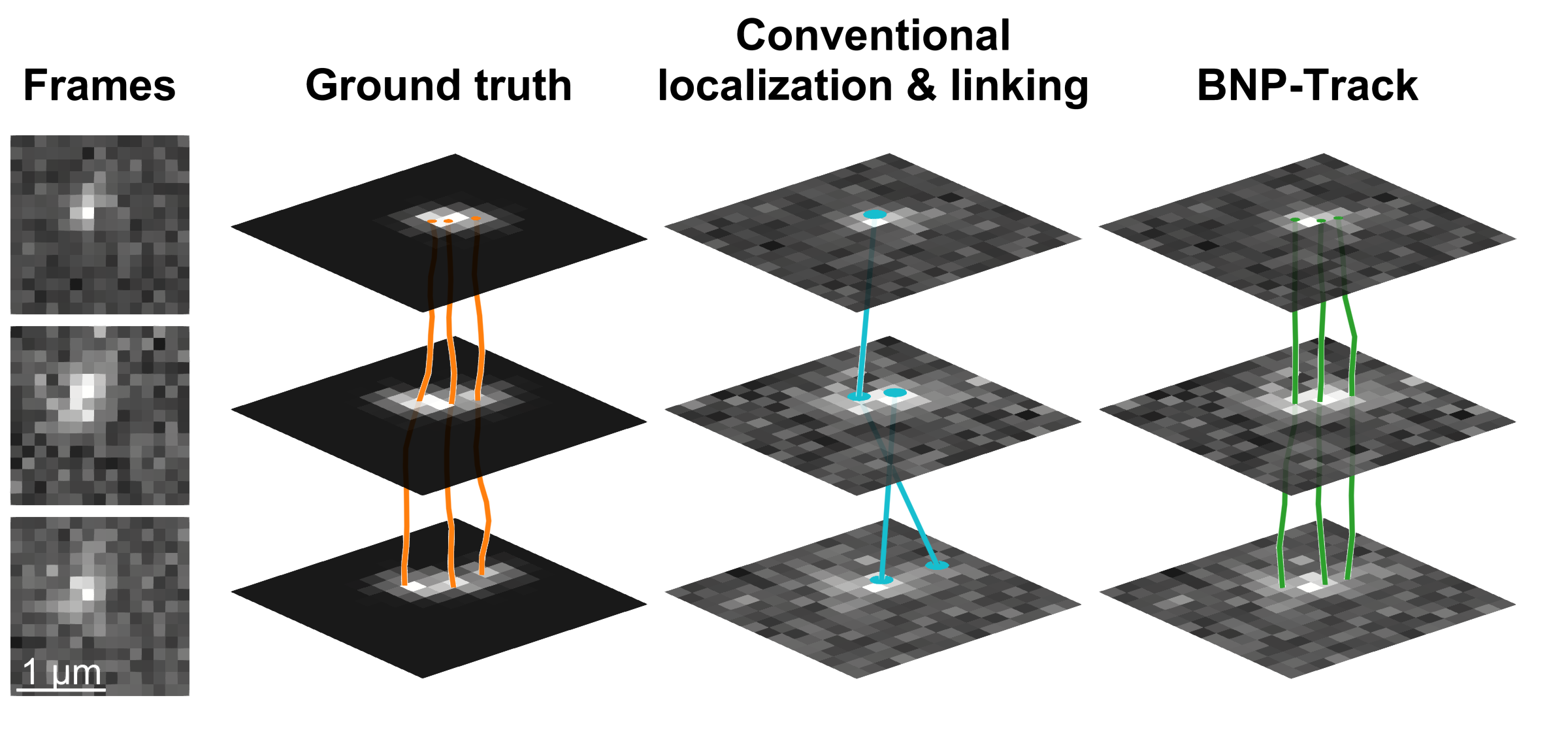}
  \caption{Demonstration of a physics-inspired SPT framework, BNP-Track\autocite{Sgouralis2024bnptrack}, using synthetic datasets modeled with EMCCD camera characteristics. Three consecutive frames illustrate significant PSF overlap and high background noise. The ground truth tracks (orange) indicate the presence of three particles. Conventional offline SPT methods, such as TrackMate\autocite{tinevez2017trackmate}, produce tracks (blue) with an incorrect number of particles and mislinked trajectories. In contrast, BNP-Track's MAP estimate (green) closely matches the ground truth, accurately resolving particle identities and paths despite the challenging imaging conditions.}
  \label{fig:four}
\end{figure}

To illustrate the effectiveness of the physics-inspired SPT approach, we highlight results from a prior analysis using BNP-Track\autocite{Sgouralis2024bnptrack}. As shown in \cref{fig:four}, BNP-Track produces particle tracks that closely align with the ground truth, significantly outperforming a widely used conventional offline SPT tool, TrackMate\autocite{tinevez2017trackmate}, particularly in scenarios with substantial PSF overlap. These results support the premise that physics-inspired SPT methods offer superior data efficiency, enabling accurate tracking with fewer detected photons---a critical advantage for leveraging high-speed detectors with reduced photon budgets per frame.

\section{Issues and outlook}
Although tracking has been a central problem across many fields for centuries, and fluorescence microscopy has advanced spatial resolution to the scale of ten nanometers or below, there is still no single SPT method that simultaneously achieves both high spatiotemporal resolution and reliable multi-particle tracking across all realistic contexts. In this perspective, we reviewed the major approaches to fluorescence SPT and identified a fundamental limitation: the widespread tendency to approximate the inherently global nature of tracking as a series of static localization steps followed by combinatorial linking. We argue that global SPT frameworks offer a compelling solution to these challenges, provided computational cost can be mitigated. By treating tracking as a fully dynamic inference problem, these methods promise to overcome current trade-offs, enabling the investigation of biological phenomena at timescales and resolutions previously inaccessible, ranging from rapid receptor-ligand interactions\autocite{kasai2014single, Yanagawa2018single, calebiro2021gprotein} to real-time monitoring of nanoparticle drug delivery\autocite{PATEL2019brief, Yao2020NanoparticleDrug}.

Furthermore, by directly analyzing single-photon data, global SPT approaches can potentially incorporate additional photon-level information, such as emission wavelength and fluorescence lifetime. This opens the door to integrating techniques like single-molecule Förster resonance energy transfer and fluorescence lifetime imaging microscopy within the same tracking framework\autocite{Xu2025Single}. As a result, global SPT approaches could enable high spatiotemporal resolution tracking and the simultaneous identification and differentiation of multiple molecular species, paving the way for advanced multi-species tracking in complex biological environments.

A central challenge for physics-inspired SPT methods is their high computational demands. Efficiently identifying the globally most probable particle tracks from a high-dimensional posterior distribution remains an open problem, particularly as the dimensionality scales with the number of particles and frames. Dealing with subtle and fundamental indeterminacies, such as approximately indistinguishable out-of-focus particles from the background, remains another computational bottleneck that ultimately demands user input.

Moreover, incorporating realistic physical models into the analysis, essential for accurately capturing the system's dynamics, often results in complex, highly structured likelihood functions that are computationally expensive to evaluate. For instance, in principle, particles should be treated as mobile during the camera's exposure period. This implies that one must integrate over all possible particle paths (a process known as marginalization in probabilistic terms) to accurately model the emission pattern a moving particle produces. Alternatively, one could demarginalize the problem by explicitly inferring short track segments for each particle within each exposure period. The resulting likelihood becomes significantly more complex in either approach, posing substantial computational challenges\autocite{Sgouralis2024bnptrack}.

While computationally demanding, this challenge is far from insurmountable, particularly in an era marked by rapid advances in parallel computing and numerical optimization. A wide range of parallelization strategies has been developed at both the processor and thread levels to accelerate high-dimensional inference tasks, including parallel MCMC, Hamiltonian Monte Carlo variants, and particle-based methods\autocite{Foreman-Mackey2013emcee, Wang2017Particle, goudie2020MultiBUGS, Heng2019Unbiased, Luengo2020survey,wang2023recent, Tian2023Recent}. These strategies enable substantial speedups, especially when tailored to exploit the structured nature of likelihoods in physics-inspired problems.

In parallel, next-generation numerical libraries are introducing mixed-precision arithmetic and randomized linear algebra techniques to improve scalability while preserving accuracy\autocite{Higham2022Mixed, murray2023randomized}. For instance, mixed-precision computation allows high-cost variables (\textit{e.g.}, posterior estimates) to be maintained in full precision, while less sensitive quantities (\textit{e.g.}, intermediate updates or gradients) are processed with lower precision, reducing memory usage and computational latency.

Hardware innovations further compound these gains. General-purpose GPUs, and increasingly, in-memory computing hardware\autocite{Lin2024Deep}, tensor processing units\autocite{Jouppi2018Motivation, Hsu2021Accelerating, Si2024carbon}, offer orders-of-magnitude improvements in throughput for numerical workloads. These platforms are becoming more accessible, making deploying high-resolution tracking algorithms on desktop systems rather than supercomputers feasible.

Collectively, these developments suggest that the computational bottlenecks historically limiting physics-inspired SPT are being eroded on multiple fronts. With the right integration of algorithmic design and hardware acceleration, particle tracking for complex biological systems is becoming increasingly attainable.

Beyond numerical acceleration, AI-augmented physical modeling offers promising yet largely untapped potential\autocite{Raissi2019Physics, Karniadakis2021Physics, Sholokhov2023Physics, Qiao2022Rationalized, Mitra2024Learning, Qiao2025neural}. The overarching goal is to synergize the computational efficiency and modeling flexibility of neural networks with the mathematical rigor of physics-based approaches. Several promising strategies have already emerged in this direction.

First, neural networks, such as normalizing flows, can be trained to approximate likelihood functions or to learn mappings from observations to posterior samples\autocite{Gabrie2022Adaptive, Winter2024Emerging, Rockova2025AI}. These hybrid frameworks retain the interpretability and structure of physics models while gaining the scalability and adaptability of deep learning, enabling fast and accurate inference in otherwise intractable settings.

Second, in scenarios where likelihoods are too complex or expensive to compute directly, neural networks can facilitate simulation-based inference\autocite{Cranmer2020frontier, Dingeldein2023Simulation}. A typical approach involves using a physical model to generate synthetic data, often far cheaper than computing explicit likelihoods, and then training a neural network to learn a surrogate likelihood or posterior from these simulations.

Beyond the current advances in AI-augmented physical modeling, the future promises even more transformative possibilities. As these two paradigms, physics-driven rigor and AI-driven adaptability, continue to converge, they may fundamentally reshape how we model, analyze, and understand complex systems. This raises compelling questions: Could embedding rigorous physical principles reduce the need for overparameterized models and massive datasets in AI? Must physically grounded models remain computationally expensive if accelerated by intelligent neural approximators? The answers to these questions are not yet known, but they define an exciting frontier. We look forward to seeing how the scientific community navigates this convergence, opening new pathways for discovery that are both principled and powerful.

\section{Conclusion}
Global, physics-inspired SPT offers a compelling framework for achieving high spatiotemporal resolution without sacrificing the ability to track multiple particles simultaneously. Its strengths are especially pronounced when paired with fast data acquisition technologies like SPDs. While the primary challenge remains its substantial computational cost, recent advances in parallelized sampling, numerical optimization, and AI-driven neural network approximators provide promising avenues for overcoming this limitation. Together, these developments are making physics-inspired SPT increasingly viable for high-throughput, complex biological applications, bringing us closer to real-time, high-fidelity insight into molecular dynamics.

\section{Acknowledgements}
S.P. acknowledges support from the NIH (R35GM148237), ARO (W911NF-23-1-0304), and NSF (Grant No. 2310610).

\section{Author declarations}
\subsection{Conflict of interest}
The authors declare no conflict of interest.

\subsection{Author contributions}
\textbf{Lance W.Q. Xu}: Conceptualization (equal); Funding acquisition (supporting); Investigation (lead); Methodology (equal); Software (lead); Visualization (lead); Writing – original draft (lead); Writing - review \& editing (equal); \textbf{Steve Pressé}: Conceptualization (equal); Funding acquisition (lead); Methodology (equal); Supervision (lead); Visualization (supporting); Writing - review \& editing (equal).

\section{Data Availability}
The data that support this study are available from the corresponding author upon reasonable request.

\printbibliography

@article{abbe1882relation,
  author    = {Abbe Hon.},
  doi       = {10.1111/j.1365-2818.1882.tb04805.x},
  issn      = {0368-3974},
  journal   = {J. R. Microsc. Soc.},
  month     = aug,
  number    = {4},
  pages     = {460--473},
  publisher = {Wiley},
  title     = {The Relation of Aperture and Power in the Microscope (continued).},
  volume    = {2},
  year      = {1882}
}

@article{Abyzov2022Conformational,
  author    = {Abyzov, Anton and Blackledge, Martin and Zweckstetter, Markus},
  doi       = {10.1021/acs.chemrev.1c00774},
  issn      = {1520-6890},
  journal   = {Chem. Rev.},
  month     = feb,
  number    = {6},
  pages     = {6719--6748},
  publisher = {American Chemical Society (ACS)},
  title     = {Conformational Dynamics of Intrinsically Disordered Proteins Regulate Biomolecular Condensate Chemistry},
  volume    = {122},
  year      = {2022}
}

@article{Adrian2005Twenty,
  author    = {Adrian, R. J.},
  doi       = {10.1007/s00348-005-0991-7},
  issn      = {1432-1114},
  journal   = {Experiments in Fluids},
  month     = jul,
  number    = {2},
  pages     = {159--169},
  publisher = {Springer Science and Business Media LLC},
  title     = {Twenty years of particle image velocimetry},
  volume    = {39},
  year      = {2005}
}

@book{Adrian2011Particle,
  address   = {Cambridge [u.a.]},
  author    = {Adrian, Ronald J. and Westerweel, Jerry},
  isbn      = {9780521440080},
  note      = {Literaturverz. S. 529 - 546},
  number    = {30},
  pagetotal = {558},
  ppn_gvk   = {638301572},
  publisher = {Cambridge Univ. Press},
  series    = {Cambridge aerospace series},
  title     = {Particle image velocimetry},
  year      = {2011}
}

@article{Ai2022Common,
  author    = {Ai, Xiaocong and Allaire, Corentin and Calace, Noemi and Czirkos, Angéla and Elsing, Markus and Ene, Irina and Farkas, Ralf and Gagnon, Louis-Guillaume and Garg, Rocky and Gessinger, Paul and Grasland, Hadrien and Gray, Heather M. and Gumpert, Christian and Hrdinka, Julia and Huth, Benjamin and Kiehn, Moritz and Klimpel, Fabian and Kolbinger, Bernadette and Krasznahorkay, Attila and Langenberg, Robert and Leggett, Charles and Mania, Georgiana and Moyse, Edward and Niermann, Joana and Osborn, Joseph D. and Rousseau, David and Salzburger, Andreas and Schlag, Bastian and Tompkins, Lauren and Yamazaki, Tomohiro and Yeo, Beomki and Zhang, Jin},
  doi       = {10.1007/s41781-021-00078-8},
  issn      = {2510-2044},
  journal   = {Computing and Software for Big Science},
  month     = apr,
  number    = {1},
  publisher = {Springer Science and Business Media LLC},
  title     = {A Common Tracking Software Project},
  volume    = {6},
  year      = {2022}
}

@article{Andersson2005Tracking,
  author    = {Andersson, S. B.},
  day       = {01},
  doi       = {10.1007/s00340-005-1801-x},
  issn      = {1432-0649},
  journal   = {Appl. Phys. B},
  month     = jun,
  number    = {7},
  pages     = {809--816},
  publisher = {Springer Science and Business Media LLC},
  title     = {Tracking a single fluorescent molecule with a confocal microscope},
  volume    = {80},
  year      = {2005}
}

@article{Andersson2008Localization,
  author    = {Andersson, Sean},
  doi       = {10.1364/oe.16.018714},
  issn      = {1094-4087},
  journal   = {Opt. Express},
  month     = oct,
  number    = {23},
  pages     = {18714},
  publisher = {Optica Publishing Group},
  title     = {Localization of a fluorescent source without numerical fitting},
  volume    = {16},
  year      = {2008}
}

@article{Andersson2011nonlinear,
  author    = {Andersson, S. B.},
  doi       = {10.1007/s00340-011-4514-3},
  issn      = {1432-0649},
  journal   = {Appl. Phys. B},
  month     = may,
  number    = {1},
  pages     = {161--173},
  publisher = {Springer Science and Business Media LLC},
  title     = {A nonlinear controller for three-dimensional tracking of a fluorescent particle in a confocal microscope},
  volume    = {104},
  year      = {2011}
}

@article{Ashley2016Tracking,
  author    = {Ashley, Trevor T. and Gan, Eric L. and Pan, Jane and Andersson, Sean B.},
  doi       = {10.1364/boe.7.003355},
  issn      = {2156-7085},
  journal   = {Biomed. Opt. Express},
  month     = aug,
  number    = {9},
  pages     = {3355},
  publisher = {Optica Publishing Group},
  title     = {Tracking single fluorescent particles in three dimensions via extremum seeking},
  volume    = {7},
  year      = {2016}
}

@article{Ashoka2023Brightness,
  author    = {Ashoka, Anila Hoskere and Aparin, Ilya O. and Reisch, Andreas and Klymchenko, Andrey S.},
  doi       = {10.1039/d2cs00464j},
  issn      = {1460-4744},
  journal   = {Chem. Soc. Rev.},
  number    = {14},
  pages     = {4525--4548},
  publisher = {Royal Society of Chemistry (RSC)},
  title     = {Brightness of fluorescent organic nanomaterials},
  volume    = {52},
  year      = {2023}
}

@article{Balasubramanian2023,
  author    = {Balasubramanian, Harikrushnan and Hobson, Chad M. and Chew, Teng-Leong and Aaron, Jesse S.},
  day       = {28},
  doi       = {10.1038/s42003-023-05468-9},
  issn      = {2399-3642},
  journal   = {Commun. Biol.},
  month     = oct,
  number    = {1},
  pages     = {1096},
  publisher = {Springer Science and Business Media LLC},
  title     = {Imagining the future of optical microscopy: everything, everywhere, all at once},
  volume    = {6},
  year      = {2023}
}

@article{Balzarotti2017nanometer,
  author  = {Francisco Balzarotti and Yvan Eilers and Klaus C. Gwosch and Arvid H. Gynnå and Volker Westphal and Fernando D. Stefani and Johan Elf and Stefan W. Hell},
  doi     = {10.1126/science.aak9913},
  journal = {Science},
  number  = {6325},
  pages   = {606--612},
  title   = {Nanometer resolution imaging and tracking of fluorescent molecules with minimal photon fluxes},
  volume  = {355},
  year    = {2017}
}

@article{BangJensen2004When,
  author    = {Bang-Jensen, Jørgen and Gutin, Gregory and Yeo, Anders},
  doi       = {10.1016/j.disopt.2004.03.007},
  issn      = {1572-5286},
  journal   = {Discrete Optim.},
  month     = nov,
  number    = {2},
  pages     = {121--127},
  publisher = {Elsevier BV},
  title     = {When the greedy algorithm fails},
  volume    = {1},
  year      = {2004}
}

@article{Barta2024proPTV,
  author    = {Barta, Robin and Bauer, Christian and Herzog, Sebastian and Schiepel, Daniel and Wagner, Claus},
  doi       = {10.1016/j.jcp.2024.113212},
  issn      = {0021-9991},
  journal   = {Journal of Computational Physics},
  month     = oct,
  pages     = {113212},
  publisher = {Elsevier BV},
  title     = {proPTV: A probability-based particle tracking velocimetry framework},
  volume    = {514},
  year      = {2024}
}

@article{Berglund2005Tracking,
  author    = {Andrew J. Berglund and Hideo Mabuchi},
  doi       = {10.1364/opex.13.008069},
  issn      = {1094-4087},
  journal   = {Opt. Express},
  month     = Oct,
  number    = {20},
  pages     = {8069--8082},
  publisher = {Optica Publishing Group},
  title     = {Tracking-FCS: Fluorescence correlation spectroscopy of individual particles},
  url       = {https://opg.optica.org/oe/abstract.cfm?URI=oe-13-20-8069},
  volume    = {13},
  year      = {2005}
}

@article{Berglund2006Performance,
  author    = {Berglund, A. J. and Mabuchi, H.},
  day       = {01},
  doi       = {10.1007/s00340-005-2111-z},
  issn      = {1432-0649},
  journal   = {Appl. Phys. B},
  month     = Apr,
  number    = {1},
  pages     = {127--133},
  publisher = {Springer Science and Business Media LLC},
  title     = {Performance bounds on single-particle tracking by fluorescence modulation},
  volume    = {83},
  year      = {2006}
}

@article{Betzig1995proposed,
  author    = {E. Betzig},
  doi       = {10.1364/ol.20.000237},
  issn      = {1539-4794},
  journal   = {Opt. Lett.},
  month     = Feb,
  number    = {3},
  pages     = {237--239},
  publisher = {Optica Publishing Group},
  title     = {Proposed method for molecular optical imaging},
  url       = {https://opg.optica.org/ol/abstract.cfm?URI=ol-20-3-237},
  volume    = {20},
  year      = {1995}
}

@article{Betzig2006imaging,
  author    = {Eric Betzig and George H. Patterson and Rachid Sougrat and O. Wolf Lindwasser and Scott Olenych and Juan S. Bonifacino and Michael W. Davidson and Jennifer Lippincott-Schwartz and Harald F. Hess},
  doi       = {10.1126/science.1127344},
  issn      = {1095-9203},
  journal   = {Science},
  month     = sep,
  number    = {5793},
  pages     = {1642--1645},
  publisher = {American Association for the Advancement of Science},
  title     = {Imaging Intracellular Fluorescent Proteins at Nanometer Resolution},
  volume    = {313},
  year      = {2006}
}

@article{Bi2021zar1,
  author    = {Bi, Guozhi and Su, Min and Li, Nan and Liang, Yu and Dang, Song and Xu, Jiachao and Hu, Meijuan and Wang, Jizong and Zou, Minxia and Deng, Yanan and Li, Qiyu and Huang, Shijia and Li, Jiejie and Chai, Jijie and He, Kangmin and Chen, Yu-hang and Zhou, Jian-Min},
  day       = {24},
  doi       = {10.1016/j.cell.2021.05.003},
  issn      = {0092-8674},
  journal   = {Cell},
  month     = jun,
  number    = {13},
  pages     = {3528--3541.e12},
  publisher = {Elsevier},
  title     = {The ZAR1 resistosome is a calcium-permeable channel triggering plant immune signaling},
  volume    = {184},
  year      = {2021}
}

@article{Bi2025sptnet,
  author       = {Bi, Cheng and Scrudders, Kevin L. and Zheng, Yue and Mahmoodi, Maryam and Low-Nam, Shalini T. and Huang, Fang},
  doi          = {10.1101/2025.02.04.636521},
  elocation-id = {2025.02.04.636521},
  journal      = {bioRxiv},
  month        = feb,
  publisher    = {Cold Spring Harbor Laboratory},
  title        = {SPTnet: a deep learning framework for end-to-end single-particle tracking and motion dynamics analysis},
  url          = {https://www.biorxiv.org/content/early/2025/02/08/2025.02.04.636521},
  year         = {2025}
}

@article{bucci20244d,
  author    = {Bucci, Andrea and Tortarolo, Giorgio and Held, Marcus Oliver and Bega, Luca and Perego, Eleonora and Castagnetti, Francesco and Bozzoni, Irene and Slenders, Eli and Vicidomini, Giuseppe},
  day       = {23},
  doi       = {10.1038/s41467-024-50512-9},
  issn      = {2041-1723},
  journal   = {Nat. Commun.},
  month     = jul,
  number    = {1},
  pages     = {6188},
  publisher = {Springer Science and Business Media LLC},
  title     = {4D Single-particle tracking with asynchronous read-out single-photon avalanche diode array detector},
  volume    = {15},
  year      = {2024}
}

@article{Bui-Thanh2021Optimality,
  author  = {Bui-Thanh, T.},
  journal = {SIAM news},
  number  = 6,
  title   = {The Optimality of Bayes' Theorem},
  url     = {https://par.nsf.gov/biblio/10288595},
  volume  = 54,
  year    = 2021
}

@article{calebiro2021gprotein,
  author    = {Calebiro, Davide and Koszegi, Zsombor and Lanoiselée, Yann and Miljus, Tamara and O'Brien, Shannon},
  doi       = {10.1152/physrev.00021.2020},
  issn      = {1522-1210},
  journal   = {Physiol. Rev.},
  month     = jul,
  number    = {3},
  pages     = {857--906},
  publisher = {American Physiological Society},
  title     = {G protein-coupled receptor-G protein interactions: a single-molecule perspective},
  volume    = {101},
  year      = {2021}
}

@article{Carsten2024MINFLUX,
  author    = {Carsten, Alexander and Failla, Antonio Virgilio and Aepfelbacher, Martin},
  doi       = {10.1111/jmi.13306},
  issn      = {1365-2818},
  journal   = {J. Microsc.},
  month     = apr,
  number    = {2},
  pages     = {219--231},
  publisher = {Wiley},
  title     = {MINFLUX nanoscopy: Visualising biological matter at the nanoscale level},
  volume    = {298},
  year      = {2024}
}

@article{Celler2013single,
  author    = {Celler, Katherine and van Wezel, Gilles P. and Willemse, Joost},
  doi       = {10.1016/j.bbrc.2013.07.016},
  issn      = {0006-291X},
  journal   = {Biochem. Biophys. Res. Commun.},
  month     = aug,
  number    = {1},
  pages     = {38--42},
  publisher = {Elsevier BV},
  title     = {Single particle tracking of dynamically localizing TatA complexes in Streptomyces coelicolor},
  url       = {https://www.sciencedirect.com/science/article/pii/S0006291X1301156X},
  volume    = {438},
  year      = {2013}
}

@article{chaudesaigues1908mouvement,
  author  = {Chaudesaigues, M},
  journal = {Comptes rendus},
  pages   = {1044--1046},
  title   = {Le mouvement brownien et la formule d'Einstein},
  volume  = {147},
  year    = {1908}
}

@article{Cheezum2001Quantitative,
  author    = {Michael K. Cheezum and William F. Walker and William H. Guilford},
  doi       = {10.1016/s0006-3495(01)75884-5},
  issn      = {0006-3495},
  journal   = {Biophys. J.},
  month     = oct,
  number    = {4},
  pages     = {2378--2388},
  publisher = {Elsevier BV},
  title     = {Quantitative Comparison of Algorithms for Tracking Single Fluorescent Particles},
  url       = {https://www.sciencedirect.com/science/article/pii/S0006349501758845},
  volume    = {81},
  year      = {2001}
}

@article{Chen2023Fluorescence,
  author    = {Chen, Pengfa and Kang, Qin and Niu, JingJing and Jing, YingYing and Zhang, Xiao and Yu, Bin and Qu, Junle and Lin, Danying},
  doi       = {10.1364/boe.485729},
  issn      = {2156-7085},
  journal   = {Biomed. Opt. Express},
  month     = mar,
  number    = {4},
  pages     = {1718},
  publisher = {Optica Publishing Group},
  title     = {Fluorescence lifetime tracking and imaging of single moving particles assisted by a low-photon-count analysis algorithm},
  volume    = {14},
  year      = {2023}
}

@article{chenouard2013multiple,
  author    = {Chenouard, N. and Bloch, I. and Olivo-Marin, J.},
  doi       = {10.1109/tpami.2013.97},
  issn      = {2160-9292},
  journal   = {IEEE Trans. Pattern Anal. Mach. Intell.},
  month     = nov,
  number    = {11},
  pages     = {2736--3750},
  publisher = {Institute of Electrical and Electronics Engineers},
  title     = {Multiple Hypothesis Tracking for Cluttered Biological Image Sequences},
  volume    = {35},
  year      = {2013}
}

@article{chenouard2014objective,
  author    = {Chenouard, Nicolas and Smal, Ihor and de Chaumont, Fabrice and Maška, Martin and Sbalzarini, Ivo F and Gong, Yuanhao and Cardinale, Janick and Carthel, Craig and Coraluppi, Stefano and Winter, Mark and Cohen, Andrew R and Godinez, William J and Rohr, Karl and Kalaidzidis, Yannis and Liang, Liang and Duncan, James and Shen, Hongying and Xu, Yingke and Magnusson, Klas E G and Jaldén, Joakim and Blau, Helen M and Paul-Gilloteaux, Perrine and Roudot, Philippe and Kervrann, Charles and Waharte, François and Tinevez, Jean-Yves and Shorte, Spencer L and Willemse, Joost and Celler, Katherine and van Wezel, Gilles P and Dan, Han-Wei and Tsai, Yuh-Show and de Solórzano, Carlos Ortiz and Olivo-Marin, Jean-Christophe and Meijering, Erik},
  doi       = {10.1038/nmeth.2808},
  issn      = {1548-7105},
  journal   = {Nat. Methods},
  month     = jan,
  number    = {3},
  pages     = {281--289},
  publisher = {Springer Science and Business Media LLC},
  title     = {Objective comparison of particle tracking methods},
  volume    = {11},
  year      = {2014}
}

@article{Clarke2022,
  author    = {Clarke, Maggie D. and Larson, Eric and Peterson, Erica R. and McCloy, Daniel R. and Bosseler, Alexis N. and Taulu, Samu},
  doi       = {10.3389/fneur.2022.827529},
  issn      = {1664-2295},
  journal   = {Front. Neurol.},
  language  = {en},
  month     = mar,
  pages     = {827529},
  publisher = {Frontiers Media SA},
  title     = {Improving Localization Accuracy of Neural Sources by Pre-processing: Demonstration with Infant MEG Data},
  volume    = {13},
  year      = {2022}
}

@article{Cox2011Bayesian,
  author    = {Cox, Susan and Rosten, Edward and Monypenny, James and Jovanovic-Talisman, Tijana and Burnette, Dylan T and Lippincott-Schwartz, Jennifer and Jones, Gareth E and Heintzmann, Rainer},
  doi       = {10.1038/nmeth.1812},
  issn      = {1548-7105},
  journal   = {Nat. Methods},
  month     = dec,
  number    = {2},
  pages     = {195--200},
  publisher = {Springer Science and Business Media LLC},
  title     = {Bayesian localization microscopy reveals nanoscale podosome dynamics},
  volume    = {9},
  year      = {2011}
}

@article{Cranmer2020frontier,
  author    = {Cranmer, Kyle and Brehmer, Johann and Louppe, Gilles},
  doi       = {10.1073/pnas.1912789117},
  issn      = {1091-6490},
  journal   = {Proc. Natl. Acad. Sci. U.S.A.},
  month     = may,
  number    = {48},
  pages     = {30055--30062},
  publisher = {Proceedings of the National Academy of Sciences},
  title     = {The frontier of simulation-based inference},
  volume    = {117},
  year      = {2020}
}

@article{CROCKER1996methods,
  author    = {John C. Crocker and David G. Grier},
  doi       = {10.1006/jcis.1996.0217},
  issn      = {0021-9797},
  journal   = {J. Colloid Interface Sci.},
  month     = apr,
  number    = {1},
  pages     = {298--310},
  publisher = {Elsevier BV},
  title     = {Methods of Digital Video Microscopy for Colloidal Studies},
  url       = {https://www.sciencedirect.com/science/article/pii/S0021979796902179},
  volume    = {179},
  year      = {1996}
}

@book{Dabiri2019Particle,
  author    = {Dabiri, Dana and Pecora, Charles},
  doi       = {10.1088/978-0-7503-2203-4},
  isbn      = {978-0-7503-2203-4},
  publisher = {IOP Publishing},
  series    = {2053-2563},
  title     = {Particle Tracking Velocimetry},
  url       = {https://dx.doi.org/10.1088/978-0-7503-2203-4},
  year      = {2019}
}

@article{Deng2009Effect,
  author    = {Deng, Yi and Shaevitz, Joshua W.},
  doi       = {10.1364/ao.48.001886},
  issn      = {1539-4522},
  journal   = {Appl. Opt.},
  month     = mar,
  number    = {10},
  pages     = {1886},
  publisher = {Optica Publishing Group},
  title     = {Effect of aberration on height calibration in three-dimensional localization-based microscopy and particle tracking},
  volume    = {48},
  year      = {2009}
}

@article{Dickson1997onoff,
  author    = {Dickson, Robert M. and Cubitt, Andrew B. and Tsien, Roger Y. and Moerner, W. E.},
  day       = {01},
  doi       = {10.1038/41048},
  issn      = {1476-4687},
  journal   = {Nature},
  month     = jul,
  number    = {6640},
  pages     = {355--358},
  publisher = {Springer Science and Business Media LLC},
  title     = {On/off blinking and switching behaviour of single molecules of green fluorescent protein},
  volume    = {388},
  year      = {1997}
}

@article{diezmann2017three,
  author    = {von Diezmann, Lexy and Shechtman, Yoav and Moerner, W. E.},
  doi       = {10.1021/acs.chemrev.6b00629},
  issn      = {1520-6890},
  journal   = {Chem. Rev.},
  month     = feb,
  number    = {11},
  pages     = {7244--7275},
  publisher = {American Chemical Society},
  title     = {Three-Dimensional Localization of Single Molecules for Super-Resolution Imaging and Single-Particle Tracking},
  volume    = {117},
  year      = {2017}
}

@article{DiMeglio2024Quantum,
  author    = {Di Meglio, Alberto and Jansen, Karl and Tavernelli, Ivano and Alexandrou, Constantia and Arunachalam, Srinivasan and Bauer, Christian W. and Borras, Kerstin and Carrazza, Stefano and Crippa, Arianna and Croft, Vincent and de Putter, Roland and Delgado, Andrea and Dunjko, Vedran and Egger, Daniel J. and Fernández-Combarro, Elias and Fuchs, Elina and Funcke, Lena and González-Cuadra, Daniel and Grossi, Michele and Halimeh, Jad C. and Holmes, Zoë and Kühn, Stefan and Lacroix, Denis and Lewis, Randy and Lucchesi, Donatella and Martinez, Miriam Lucio and Meloni, Federico and Mezzacapo, Antonio and Montangero, Simone and Nagano, Lento and Pascuzzi, Vincent R. and Radescu, Voica and Ortega, Enrique Rico and Roggero, Alessandro and Schuhmacher, Julian and Seixas, Joao and Silvi, Pietro and Spentzouris, Panagiotis and Tacchino, Francesco and Temme, Kristan and Terashi, Koji and Tura, Jordi and Tüysüz, Cenk and Vallecorsa, Sofia and Wiese, Uwe-Jens and Yoo, Shinjae and Zhang, Jinglei},
  doi       = {10.1103/prxquantum.5.037001},
  issn      = {2691-3399},
  journal   = {PRX Quantum},
  month     = aug,
  number    = {3},
  pages     = {037001},
  publisher = {American Physical Society (APS)},
  title     = {Quantum Computing for High-Energy Physics: State of the Art and Challenges},
  volume    = {5},
  year      = {2024}
}

@article{Dingeldein2023Simulation,
  author    = {Dingeldein, Lars and Cossio, Pilar and Covino, Roberto},
  doi       = {10.1088/2632-2153/acc8b8},
  issn      = {2632-2153},
  journal   = {Mach. Learn.: Sci. Technol.},
  month     = apr,
  number    = {2},
  pages     = {025009},
  publisher = {IOP Publishing},
  title     = {Simulation-based inference of single-molecule force spectroscopy},
  volume    = {4},
  year      = {2023}
}

@article{Einstein1905Ueber,
  author    = {Einstein, A.},
  doi       = {10.1002/andp.19053220806},
  issn      = {1521-3889},
  journal   = {Ann. Phys.},
  month     = jan,
  number    = {8},
  pages     = {549--560},
  publisher = {Wiley},
  title     = {Über die von der molekularkinetischen Theorie der Wärme geforderte Bewegung von in ruhenden Flüssigkeiten suspendierten Teilchen},
  volume    = {322},
  year      = {1905}
}

@article{Elf2007Probing,
  author    = {Elf, Johan and Li, Gene-Wei and Xie, X. Sunney},
  doi       = {10.1126/science.1141967},
  issn      = {1095-9203},
  journal   = {Science},
  month     = may,
  number    = {5828},
  pages     = {1191--1194},
  publisher = {American Association for the Advancement of Science (AAAS)},
  title     = {Probing Transcription Factor Dynamics at the Single-Molecule Level in a Living Cell},
  volume    = {316},
  year      = {2007}
}

@article{Elson1974Fluorescence,
  author    = {Elson, Elliot L. and Magde, Douglas},
  doi       = {10.1002/bip.1974.360130102},
  issn      = {1097-0282},
  journal   = {Biopolymers},
  month     = jan,
  number    = {1},
  pages     = {1--27},
  publisher = {Wiley},
  title     = {Fluorescence correlation spectroscopy. I. Conceptual basis and theory},
  volume    = {13},
  year      = {1974}
}

@article{Elson2011Fluorescence,
  author    = {Elliot L. Elson},
  doi       = {10.1016/j.bpj.2011.11.012},
  issn      = {0006-3495},
  journal   = {Biophys. J.},
  month     = dec,
  number    = {12},
  pages     = {2855--2870},
  publisher = {Elsevier BV},
  title     = {Fluorescence Correlation Spectroscopy: Past, Present, Future},
  url       = {https://www.sciencedirect.com/science/article/pii/S0006349511013294},
  volume    = {101},
  year      = {2011}
}

@article{Emami2021Computerized,
  author    = {Emami, Neda and Sedaei, Zahra and Ferdousi, Reza},
  doi       = {10.1016/j.visinf.2020.11.003},
  issn      = {2468-502X},
  journal   = {Visual Informatics},
  month     = mar,
  number    = {1},
  pages     = {1--13},
  publisher = {Elsevier BV},
  title     = {Computerized cell tracking: Current methods, tools and challenges},
  volume    = {5},
  year      = {2021}
}

@article{Enderlein2000tracking,
  author    = {Enderlein, J.},
  day       = {01},
  doi       = {10.1007/s003400000409},
  issn      = {1432-0649},
  journal   = {Appl. Phys. B},
  month     = nov,
  number    = {5},
  pages     = {773--777},
  publisher = {Springer Science and Business Media LLC},
  title     = {Tracking of fluorescent molecules diffusing within membranes},
  volume    = {71},
  year      = {2000}
}

@article{Ershov2022TrackMate,
  author    = {Ershov, Dmitry and Phan, Minh-Son and Pylvänäinen, Joanna W. and Rigaud, Stéphane U. and Le Blanc, Laure and Charles-Orszag, Arthur and Conway, James R. W. and Laine, Romain F. and Roy, Nathan H. and Bonazzi, Daria and Duménil, Guillaume and Jacquemet, Guillaume and Tinevez, Jean-Yves},
  day       = {01},
  doi       = {10.1038/s41592-022-01507-1},
  issn      = {1548-7105},
  journal   = {Nat. Methods},
  month     = Jul,
  number    = {7},
  pages     = {829--832},
  publisher = {Springer Science and Business Media LLC},
  title     = {TrackMate 7: integrating state-of-the-art segmentation algorithms into tracking pipelines},
  volume    = {19},
  year      = {2022}
}

@article{Faraz2022Deep,
  author    = {Faraz, Khuram and Grenier, Thomas and Ducottet, Christophe and Epicier, Thierry},
  day       = {15},
  doi       = {10.1038/s41598-022-06308-2},
  issn      = {2045-2322},
  journal   = {Sci. Rep.},
  month     = feb,
  number    = {1},
  pages     = {2484},
  publisher = {Springer Science and Business Media LLC},
  title     = {Deep learning detection of nanoparticles and multiple object tracking of their dynamic evolution during in situ ETEM studies},
  volume    = {12},
  year      = {2022}
}

@article{Fazel2019Bayesian,
  author    = {Fazel, Mohamadreza and Wester, Michael J. and Mazloom-Farsibaf, Hanieh and Meddens, Marjolein B. M. and Eklund, Alexandra S. and Schlichthaerle, Thomas and Schueder, Florian and Jungmann, Ralf and Lidke, Keith A.},
  doi       = {10.1038/s41598-019-50232-x},
  issn      = {2045-2322},
  journal   = {Sci. Rep.},
  month     = sep,
  number    = {1},
  publisher = {Springer Science and Business Media LLC},
  title     = {Bayesian Multiple Emitter Fitting using Reversible Jump Markov Chain Monte Carlo},
  volume    = {9},
  year      = {2019}
}

@article{fazel2024fluorescence,
  author    = {Fazel, Mohamadreza and Grussmayer, Kristin S. and Ferdman, Boris and Radenovic, Aleksandra and Shechtman, Yoav and Enderlein, Jörg and Pressé, Steve},
  doi       = {10.1103/revmodphys.96.025003},
  issn      = {1539-0756},
  issue     = {2},
  journal   = {Rev. Mod. Phys.},
  month     = jun,
  number    = {2},
  numpages  = {74},
  pages     = {025003},
  publisher = {American Physical Society},
  title     = {Fluorescence microscopy: A statistics-optics perspective},
  volume    = {96},
  year      = {2024}
}

@article{Fazel2025Simultaneous,
  author    = {Fazel, Mohamadreza and Hoseini, Reza and Mahmoodi, Maryam and Xu, Lance W. Q. and Saurabh, Ayush and Kilic, Zeliha and Antolin, Julian and Scrudders, Kevin L. and Shepherd, Douglas and Low-Nam, Shalini T. and Huang, Fang and Pressé, Steve},
  doi       = {10.1101/2025.05.02.651986},
  month     = may,
  publisher = {Cold Spring Harbor Laboratory},
  title     = {Simultaneous particle tracking, phase retrieval and point spread function reconstruction},
  year      = {2025}
}

@article{Feng2007Accurate,
  author    = {Feng, Y. and Goree, J. and Liu, Bin},
  doi       = {10.1063/1.2735920},
  issn      = {1089-7623},
  journal   = {Rev. Sci. Instrum.},
  month     = may,
  number    = {5},
  pages     = {053704},
  publisher = {AIP Publishing},
  title     = {Accurate particle position measurement from images},
  volume    = {78},
  year      = {2007}
}

@article{fernandez-suarez_2008,
  author    = {Fern{\'a}ndez-Su{\'a}rez, M and Ting, AY},
  doi       = {10.1038/nrm2531},
  issn      = {1471-0080},
  journal   = {Nat. Rev. Mol. Cell Biol.},
  month     = nov,
  number    = {12},
  pages     = {929--943},
  publisher = {Springer Science and Business Media LLC},
  title     = {Fluorescent probes for super-resolution imaging in living cells},
  volume    = {9},
  year      = {2008}
}

@article{Flors2007Stroboscopic,
  author    = {Flors, Cristina and Hotta, Jun-ichi and Uji-i, Hiroshi and Dedecker, Peter and Ando, Ryoko and Mizuno, Hideaki and Miyawaki, Atsushi and Hofkens, Johan},
  doi       = {10.1021/ja074704l},
  issn      = {1520-5126},
  journal   = {Journal of the American Chemical Society},
  month     = oct,
  number    = {45},
  pages     = {13970--13977},
  publisher = {American Chemical Society (ACS)},
  title     = {A Stroboscopic Approach for Fast Photoactivation-Localization Microscopy with Dronpa Mutants},
  volume    = {129},
  year      = {2007}
}

@article{Foreman-Mackey2013emcee,
  author    = {Daniel Foreman-Mackey and David W. Hogg and Dustin Lang and Jonathan Goodman},
  doi       = {10.1086/670067},
  issn      = {1538-3873},
  journal   = {Publ. Astron. Soc. Pac.},
  month     = mar,
  number    = {925},
  pages     = {306--312},
  publisher = {IOP Publishing},
  title     = {emcee: The MCMC Hammer},
  volume    = {125},
  year      = {2013}
}

@article{Gabrie2022Adaptive,
  author    = {Marylou Gabrié and Grant M. Rotskoff and Eric Vanden-Eijnden},
  doi       = {10.1073/pnas.2109420119},
  issn      = {1091-6490},
  journal   = {Proc. Natl. Acad. Sci. U.S.A.},
  month     = mar,
  number    = {10},
  pages     = {e2109420119},
  publisher = {Proceedings of the National Academy of Sciences},
  title     = {Adaptive Monte Carlo augmented with normalizing flows},
  volume    = {119},
  year      = {2022}
}

@article{Gallatin2012Optimal,
  author    = {Gregg M. Gallatin and Andrew J. Berglund},
  doi       = {10.1364/oe.20.016381},
  issn      = {1094-4087},
  journal   = {Opt. Express},
  month     = jul,
  number    = {15},
  pages     = {16381--16393},
  publisher = {Optica Publishing Group},
  title     = {Optimal laser scan path for localizing a fluorescent particle in two or three dimensions},
  url       = {https://opg.optica.org/oe/abstract.cfm?URI=oe-20-15-16381},
  volume    = {20},
  year      = {2012}
}

@article{Galvanetto2023extreme,
  author    = {Galvanetto, Nicola and Ivanović, Miloš T. and Chowdhury, Aritra and Sottini, Andrea and Nüesch, Mark F. and Nettels, Daniel and Best, Robert B. and Schuler, Benjamin},
  day       = {01},
  doi       = {10.1038/s41586-023-06329-5},
  issn      = {1476-4687},
  journal   = {Nature},
  month     = jul,
  number    = {7971},
  pages     = {876--883},
  publisher = {Springer Science and Business Media LLC},
  title     = {Extreme dynamics in a biomolecular condensate},
  volume    = {619},
  year      = {2023}
}

@article{geerts1987nanovid,
  author    = {Geerts, Hugo and De Brabander, M and Nuydens, Ronny and Geuens, Staf and Moeremans, Mark and De Mey, J and Hollenbeck, Peter},
  doi       = {10.1016/s0006-3495(87)83271-x},
  issn      = {0006-3495},
  journal   = {Biophys. J.},
  month     = nov,
  number    = {5},
  pages     = {775--782},
  publisher = {Elsevier},
  title     = {Nanovid tracking: a new automatic method for the study of mobility in living cells based on colloidal gold and video microscopy},
  volume    = {52},
  year      = {1987}
}

@article{goudie2020MultiBUGS,
  author  = {Goudie, Robert J. B. and Turner, Rebecca M. and De Angelis, Daniela and Thomas, Andrew},
  doi     = {10.18637/jss.v095.i07},
  emsid   = {EMS85408},
  journal = {J. Stat. Softw.},
  number  = {7},
  pages   = {1},
  title   = {MultiBUGS: A Parallel Implementation of the BUGS Modeling Framework for Faster Bayesian Inference},
  volume  = {95},
  year    = {2020}
}

@article{Grimm2021Caveat,
  author    = {Grimm, Jonathan B. and Lavis, Luke D.},
  doi       = {10.1038/s41592-021-01338-6},
  issn      = {1548-7105},
  journal   = {Nat. Methods},
  month     = dec,
  number    = {2},
  pages     = {149--158},
  publisher = {Springer Science and Business Media LLC},
  title     = {Caveat fluorophore: an insiders' guide to small-molecule fluorescent labels},
  volume    = {19},
  year      = {2021}
}

@article{Gustafsson2000Surpassing,
  author    = {Gustafsson, M. G. L.},
  doi       = {10.1046/j.1365-2818.2000.00710.x},
  issn      = {1365-2818},
  journal   = {J. Microsc.},
  month     = may,
  number    = {2},
  pages     = {82--87},
  publisher = {Wiley},
  title     = {Surpassing the lateral resolution limit by a factor of two using structured illumination microscopy},
  volume    = {198},
  year      = {2000}
}

@article{Gwosch2020MINFLUX,
  author    = {Gwosch, Klaus C. and Pape, Jasmin K. and Balzarotti, Francisco and Hoess, Philipp and Ellenberg, Jan and Ries, Jonas and Hell, Stefan W.},
  day       = {01},
  doi       = {10.1038/s41592-019-0688-0},
  issn      = {1548-7105},
  journal   = {Nat. Methods},
  month     = feb,
  number    = {2},
  pages     = {217--224},
  publisher = {Springer Science and Business Media LLC},
  title     = {MINFLUX nanoscopy delivers 3D multicolor nanometer resolution in cells},
  volume    = {17},
  year      = {2020}
}

@inproceedings{Gyongy2018highspeed,
  author       = {Istvan Gyongy and Amy Davies and Allende Miguelez Crespo and Andrew Green and Neale A. W. Dutton and Rory R. Duncan and Colin Rickman and Robert K. Henderson and Paul A. Dalgarno},
  booktitle    = {High-Speed Biomedical Imaging and Spectroscopy III: Toward Big Data Instrumentation and Management},
  doi          = {10.1117/12.2290199},
  editor       = {Kevin K. Tsia and Keisuke Goda},
  month        = feb,
  organization = {International Society for Optics and Photonics},
  pages        = {105050A},
  publisher    = {SPIE},
  title        = {High-speed particle tracking in microscopy using SPAD image sensors},
  year         = {2018}
}

@article{Heckert2022Recovering,
  author    = {Heckert, Alec and Dahal, Liza and Tjian, Robert and Darzacq, Xavier},
  doi       = {10.7554/elife.70169},
  issn      = {2050-084X},
  journal   = {eLife},
  month     = sep,
  publisher = {eLife Sciences Publications, Ltd},
  title     = {Recovering mixtures of fast-diffusing states from short single-particle trajectories},
  volume    = {11},
  year      = {2022}
}

@article{Heilemann2008Subdiffraction,
  author    = {Heilemann, Mike and van de Linde, Sebastian and Schüttpelz, Mark and Kasper, Robert and Seefeldt, Britta and Mukherjee, Anindita and Tinnefeld, Philip and Sauer, Markus},
  doi       = {10.1002/anie.200802376},
  issn      = {1521-3773},
  journal   = {Angew. Chem. - Int. Ed.},
  month     = jul,
  number    = {33},
  pages     = {6172--6176},
  publisher = {Wiley},
  title     = {Subdiffraction-Resolution Fluorescence Imaging with Conventional Fluorescent Probes},
  volume    = {47},
  year      = {2008}
}

@article{Heintzmann2002Saturated,
  author    = {Rainer Heintzmann and Thomas M. Jovin and Christoph Cremer},
  doi       = {10.1364/josaa.19.001599},
  issn      = {1520-8532},
  journal   = {J. Opt. Soc. Am. A},
  month     = aug,
  number    = {8},
  pages     = {1599--1609},
  publisher = {Optica Publishing Group},
  title     = {Saturated patterned excitation microscopy—a concept for optical resolution improvement},
  url       = {https://opg.optica.org/josaa/abstract.cfm?URI=josaa-19-8-1599},
  volume    = {19},
  year      = {2002}
}

@article{Hell1994breaking,
  author    = {Stefan W. Hell and Jan Wichmann},
  doi       = {10.1364/ol.19.000780},
  issn      = {1539-4794},
  journal   = {Opt. Lett.},
  month     = jun,
  number    = {11},
  pages     = {780--782},
  publisher = {Optica Publishing Group},
  title     = {Breaking the diffraction resolution limit by stimulated emission: stimulated-emission-depletion fluorescence microscopy},
  url       = {https://opg.optica.org/ol/abstract.cfm?URI=ol-19-11-780},
  volume    = {19},
  year      = {1994}
}

@article{Heng2019Unbiased,
  author    = {Heng, J and Jacob, P E},
  doi       = {10.1093/biomet/asy074},
  issn      = {1464-3510},
  journal   = {Biometrika},
  month     = feb,
  number    = {2},
  pages     = {287--302},
  publisher = {Oxford University Press (OUP)},
  title     = {Unbiased Hamiltonian Monte Carlo with couplings},
  volume    = {106},
  year      = {2019}
}

@article{HESS2006ultra,
  author    = {Samuel T. Hess and Thanu P.K. Girirajan and Michael D. Mason},
  doi       = {10.1529/biophysj.106.091116},
  issn      = {0006-3495},
  journal   = {Biophys. J.},
  month     = dec,
  number    = {11},
  pages     = {4258--4272},
  publisher = {Elsevier},
  title     = {Ultra-High Resolution Imaging by Fluorescence Photoactivation Localization Microscopy},
  url       = {https://www.sciencedirect.com/science/article/pii/S0006349506721403},
  volume    = {91},
  year      = {2006}
}

@article{Higham2022Mixed,
  author    = {Higham, Nicholas J. and Mary, Theo},
  doi       = {10.1017/s0962492922000022},
  issn      = {1474-0508},
  journal   = {Acta Numer.},
  month     = may,
  pages     = {347--414},
  publisher = {Cambridge University Press},
  title     = {Mixed precision algorithms in numerical linear algebra},
  volume    = {31},
  year      = {2022}
}

@article{Holden2011,
  author    = {Holden, Seamus J and Uphoff, Stephan and Kapanidis, Achillefs N},
  doi       = {10.1038/nmeth0411-279},
  issn      = {1548-7105},
  journal   = {Nat. Methods},
  month     = mar,
  number    = {4},
  pages     = {279--280},
  publisher = {Springer Science and Business Media LLC},
  title     = {DAOSTORM: an algorithm for high- density super-resolution microscopy},
  volume    = {8},
  year      = {2011}
}

@article{Hou2019Adaptive,
  author    = {Hou, Shangguo and Welsher, Kevin},
  doi       = {10.1002/smll.201903039},
  issn      = {1613-6829},
  journal   = {Small},
  month     = sep,
  number    = {44},
  pages     = {1903039},
  publisher = {Wiley},
  title     = {An Adaptive Real‐Time 3D Single Particle Tracking Method for Monitoring Viral First Contacts},
  volume    = {15},
  year      = {2019}
}

@article{Hou2020Real,
  author    = {Hou, Shangguo and Exell, Jack and Welsher, Kevin},
  doi       = {10.1038/s41467-020-17444-6},
  issn      = {2041-1723},
  journal   = {Nat. Commun.},
  month     = jul,
  number    = {1},
  publisher = {Springer Science and Business Media LLC},
  title     = {Real-time 3D single molecule tracking},
  volume    = {11},
  year      = {2020}
}

@inproceedings{Hsu2021Accelerating,
  author     = {Hsu, Kuan-Chieh and Tseng, Hung-Wei},
  booktitle  = {Proceedings of the International Conference for High Performance Computing, Networking, Storage and Analysis},
  collection = {SC '21},
  doi        = {10.1145/3458817.3476177},
  month      = nov,
  publisher  = {ACM},
  series     = {SC '21},
  title      = {Accelerating applications using edge tensor processing units},
  year       = {2021}
}

@article{Huang2011Simultaneous,
  author    = {Huang, Fang and Schwartz, Samantha L. and Byars, Jason M. and Lidke, Keith A.},
  doi       = {10.1364/boe.2.001377},
  issn      = {2156-7085},
  journal   = {Biomed. Opt. Express},
  month     = apr,
  number    = {5},
  pages     = {1377},
  publisher = {Optica Publishing Group},
  title     = {Simultaneous multiple-emitter fitting for single molecule super-resolution imaging},
  volume    = {2},
  year      = {2011}
}

@article{jaqaman2008robust,
  author    = {Jaqaman, Khuloud and Loerke, Dinah and Mettlen, Marcel and Kuwata, Hirotaka and Grinstein, Sergio and Schmid, Sandra L and Danuser, Gaudenz},
  doi       = {10.1038/nmeth.1237},
  issn      = {1548-7105},
  journal   = {Nat. Methods},
  month     = jul,
  number    = {8},
  pages     = {695--702},
  publisher = {Springer Science and Business Media LLC},
  title     = {Robust single-particle tracking in live-cell time-lapse sequences},
  volume    = {5},
  year      = {2008}
}

@article{Jazani2019Alternative,
  author    = {Jazani, Sina and Sgouralis, Ioannis and Shafraz, Omer M. and Levitus, Marcia and Sivasankar, Sanjeevi and Pressé, Steve},
  doi       = {10.1038/s41467-019-11574-2},
  issn      = {2041-1723},
  journal   = {Nat. Commun.},
  month     = aug,
  number    = {1},
  pages     = {3662},
  publisher = {Springer Science and Business Media LLC},
  title     = {An alternative framework for fluorescence correlation spectroscopy},
  volume    = {10},
  year      = {2019}
}

@article{jazani2022computational,
  author    = {Jazani, Sina and Xu, Lance W. Q. and Sgouralis, Ioannis and Shepherd, Douglas P. and Pressé, Steve},
  doi       = {10.1021/acsphotonics.2c00614},
  issn      = {2330-4022},
  journal   = {ACS Photonics},
  month     = jul,
  number    = {7},
  pages     = {2489--2498},
  publisher = {American Chemical Society},
  title     = {Computational Proposal for Tracking Multiple Molecules in a Multifocus Confocal Setup},
  volume    = {9},
  year      = {2022}
}

@article{Ji2017Adaptive,
  author    = {Ji, Na},
  doi       = {10.1038/nmeth.4218},
  issn      = {1548-7105},
  journal   = {Nature Methods},
  month     = apr,
  number    = {4},
  pages     = {374--380},
  publisher = {Springer Science and Business Media LLC},
  title     = {Adaptive optical fluorescence microscopy},
  volume    = {14},
  year      = {2017}
}

@article{Johnson2022Capturing,
  author    = {Johnson, Courtney and Exell, Jack and Lin, Yuxin and Aguilar, Jonathan and Welsher, Kevin D.},
  day       = {01},
  doi       = {10.1038/s41592-022-01672-3},
  issn      = {1548-7105},
  journal   = {Nat. Methods},
  month     = dec,
  number    = {12},
  pages     = {1642--1652},
  publisher = {Springer Science and Business Media LLC},
  title     = {Capturing the start point of the virus–cell interaction with high-speed 3D single-virus tracking},
  volume    = {19},
  year      = {2022}
}

@article{Jouppi2018Motivation,
  author    = {Jouppi, Norman and Young, Cliff and Patil, Nishant and Patterson, David},
  doi       = {10.1109/mm.2018.032271057},
  issn      = {1937-4143},
  journal   = {IEEE Micro},
  month     = may,
  number    = {3},
  pages     = {10--19},
  publisher = {Institute of Electrical and Electronics Engineers (IEEE)},
  title     = {Motivation for and Evaluation of the First Tensor Processing Unit},
  volume    = {38},
  year      = {2018}
}

@article{Jungmann2010Single,
  author    = {Jungmann, Ralf and Steinhauer, Christian and Scheible, Max and Kuzyk, Anton and Tinnefeld, Philip and Simmel, Friedrich C.},
  doi       = {10.1021/nl103427w},
  issn      = {1530-6992},
  journal   = {Nano Lett.},
  month     = oct,
  number    = {11},
  pages     = {4756--4761},
  publisher = {American Chemical Society},
  title     = {Single-Molecule Kinetics and Super-Resolution Microscopy by Fluorescence Imaging of Transient Binding on DNA Origami},
  volume    = {10},
  year      = {2010}
}

@article{Jungmann2016Quantitative,
  author    = {Jungmann, Ralf and Avendaño, Maier S and Dai, Mingjie and Woehrstein, Johannes B and Agasti, Sarit S and Feiger, Zachary and Rodal, Avital and Yin, Peng},
  doi       = {10.1038/nmeth.3804},
  issn      = {1548-7105},
  journal   = {Nat. Methods},
  month     = mar,
  number    = {5},
  pages     = {439--442},
  publisher = {Springer Science and Business Media LLC},
  title     = {Quantitative super-resolution imaging with qPAINT},
  volume    = {13},
  year      = {2016}
}

@article{Kadam2024Object,
  author    = {Kadam, Pushkar and Fang, Gu and Zou, Ju Jia},
  doi       = {10.3390/computers13060136},
  issn      = {2073-431X},
  journal   = {Computers},
  month     = may,
  number    = {6},
  pages     = {136},
  publisher = {MDPI AG},
  title     = {Object Tracking Using Computer Vision: A Review},
  volume    = {13},
  year      = {2024}
}

@article{Kappler1931Versuche,
  author    = {Kappler, Eugen},
  doi       = {10.1002/andp.19314030208},
  issn      = {1521-3889},
  journal   = {Ann. Phys.},
  month     = jan,
  number    = {2},
  pages     = {233--256},
  publisher = {Wiley},
  title     = {Versuche zur Messung der Avogadro-Loschmidtschen Zahl aus der Brownschen Bewegung einer Drehwaage},
  volume    = {403},
  year      = {1931}
}

@article{Karniadakis2021Physics,
  author    = {Karniadakis, George Em and Kevrekidis, Ioannis G. and Lu, Lu and Perdikaris, Paris and Wang, Sifan and Yang, Liu},
  doi       = {10.1038/s42254-021-00314-5},
  issn      = {2522-5820},
  journal   = {Nat. Rev. Phys.},
  month     = may,
  number    = {6},
  pages     = {422--440},
  publisher = {Springer Science and Business Media LLC},
  title     = {Physics-informed machine learning},
  volume    = {3},
  year      = {2021}
}

@article{kasai2014single,
  author    = {Rinshi S Kasai and Akihiro Kusumi},
  doi       = {10.1016/j.ceb.2013.11.008},
  issn      = {0955-0674},
  journal   = {Curr. Opin. Cell Biol.},
  month     = apr,
  pages     = {78--86},
  publisher = {Elsevier BV},
  title     = {Single-molecule imaging revealed dynamic GPCR dimerization},
  url       = {https://www.sciencedirect.com/science/article/pii/S0955067413001841},
  volume    = {27},
  year      = {2014}
}

@article{Katayama2009Real,
  author    = {Katayama, Yoshihiko and Burkacky, Ondrej and Meyer, Martin and Bräuchle, Christoph and Gratton, Enrico and Lamb, Don C.},
  doi       = {10.1002/cphc.200900436},
  issn      = {1439-7641},
  journal   = {ChemPhysChem},
  month     = sep,
  number    = {14},
  pages     = {2458--2464},
  publisher = {Wiley},
  title     = {Real‐Time Nanomicroscopy via Three‐Dimensional Single‐Particle Tracking},
  volume    = {10},
  year      = {2009}
}

@book{Kay1993fundamentals,
  address   = {Upper Saddle River, NJ},
  author    = {Kay, Steven M.},
  edition   = {20. pr.},
  editor    = {Steven M. Kay},
  isbn      = {0133457117},
  number    = {Estimation theory},
  pagetotal = {595},
  place     = {Englewood Cliffs},
  ppn_gvk   = {1453731350},
  publisher = {Prentice Hall PTR},
  title     = {Fundamentals of Statistical Signal Processing},
  volume    = {1},
  year      = {2013}
}

@book{Kepler2015Astronomia,
  author    = {Kepler, Johannes and Donahue, William H.},
  isbn      = {9781888009477},
  publisher = {Tradeselect Limited},
  title     = {Astronomia Nova},
  year      = {2015}
}

@article{Klar2000Fluorescence,
  author  = {Thomas A. Klar and Stefan Jakobs and Marcus Dyba and Alexander Egner and Stefan W. Hell},
  doi     = {10.1073/pnas.97.15.8206},
  journal = {Proc. Natl. Acad. Sci. U.S.A},
  number  = {15},
  pages   = {8206--8210},
  title   = {Fluorescence microscopy with diffraction resolution barrier broken by stimulated emission},
  volume  = {97},
  year    = {2000}
}

@article{lee2017unraveling,
  author    = {Lee, Antony and Tsekouras, Konstantinos and Calderon, Christopher and Bustamante, Carlos and Pressé, Steve},
  doi       = {10.1021/acs.chemrev.6b00729},
  issn      = {1520-6890},
  journal   = {Chem. Rev.},
  month     = apr,
  number    = {11},
  pages     = {7276--7330},
  publisher = {American Chemical Society},
  title     = {Unraveling the Thousand Word Picture: An Introduction to Super-Resolution Data Analysis},
  volume    = {117},
  year      = {2017}
}

@article{Lelek2021single,
  author    = {Lelek, Mickaël and Gyparaki, Melina T. and Beliu, Gerti and Schueder, Florian and Griffié, Juliette and Manley, Suliana and Jungmann, Ralf and Sauer, Markus and Lakadamyali, Melike and Zimmer, Christophe},
  day       = {03},
  doi       = {10.1038/s43586-021-00038-x},
  issn      = {2662-8449},
  journal   = {Nat. Rev. Methods Primers},
  month     = jun,
  number    = {1},
  pages     = {39},
  publisher = {Springer Science and Business Media LLC},
  title     = {Single-molecule localization microscopy},
  volume    = {1},
  year      = {2021}
}

@article{Lemmer2008spdm,
  author    = {Lemmer, P. and Gunkel, M. and Baddeley, D. and Kaufmann, R. and Urich, A. and Weiland, Y. and Reymann, J. and Müller, P. and Hausmann, M. and Cremer, C.},
  day       = {01},
  doi       = {10.1007/s00340-008-3152-x},
  issn      = {1432-0649},
  journal   = {Appl. Phys. B},
  month     = oct,
  number    = {1},
  pages     = {1--12},
  publisher = {Springer Science and Business Media LLC},
  title     = {SPDM: light microscopy with single-molecule resolution at the nanoscale},
  volume    = {93},
  year      = {2008}
}

@article{Levi20053,
  author    = {Levi, Valeria and Ruan, QiaoQiao and Gratton, Enrico},
  doi       = {10.1529/biophysj.104.044230},
  issn      = {0006-3495},
  journal   = {Biophys. J.},
  month     = apr,
  number    = {4},
  pages     = {2919--2928},
  publisher = {Elsevier BV},
  title     = {3-D Particle Tracking in a Two-Photon Microscope: Application to the Study of Molecular Dynamics in Cells},
  volume    = {88},
  year      = {2005}
}

@article{Li2019Divide,
  author    = {Li, Luchang and Xin, Bo and Kuang, Weibing and Zhou, Zhiwei and Huang, Zhen-Li},
  doi       = {10.1364/oe.27.021029},
  issn      = {1094-4087},
  journal   = {Opt. Express},
  month     = jul,
  number    = {15},
  pages     = {21029},
  publisher = {Optica Publishing Group},
  title     = {Divide and conquer: real-time maximum likelihood fitting of multiple emitters for super-resolution localization microscopy},
  volume    = {27},
  year      = {2019}
}

@article{Lidke2005Superresolution,
  author    = {Keith A. Lidke and Bernd Rieger and Thomas M. Jovin and Rainer Heintzmann},
  doi       = {10.1364/opex.13.007052},
  issn      = {1094-4087},
  journal   = {Opt. Express},
  month     = Sep,
  number    = {18},
  pages     = {7052--7062},
  publisher = {Optica Publishing Group},
  title     = {Superresolution by localization of quantum dots using blinking statistics},
  url       = {https://opg.optica.org/oe/abstract.cfm?URI=oe-13-18-7052},
  volume    = {13},
  year      = {2005}
}

@article{Lin2024Deep,
  author    = {Lin, Yudeng and Gao, Bin and Tang, Jianshi and Zhang, Qingtian and Qian, He and Wu, Huaqiang},
  doi       = {10.1038/s43588-024-00744-y},
  issn      = {2662-8457},
  journal   = {Nat. Comput. Sci.},
  month     = dec,
  number    = {1},
  pages     = {27--36},
  publisher = {Springer Science and Business Media LLC},
  title     = {Deep Bayesian active learning using in-memory computing hardware},
  volume    = {5},
  year      = {2024}
}

@article{Lionnet2021SMT,
  author    = {Lionnet, Timothee and Wu, Christine},
  doi       = {10.1016/j.gde.2020.12.001},
  issn      = {0959-437X},
  journal   = {Curr. Opin. Genet. Dev.},
  month     = apr,
  pages     = {94--102},
  publisher = {Elsevier BV},
  title     = {Single-molecule tracking of transcription protein dynamics in living cells: seeing is believing, but what are we seeing?},
  volume    = {67},
  year      = {2021}
}

@article{Liu2024Universal,
  author    = {Liu, Sheng and Chen, Jianwei and Hellgoth, Jonas and Müller, Lucas-Raphael and Ferdman, Boris and Karras, Christian and Xiao, Dafei and Lidke, Keith A. and Heintzmann, Rainer and Shechtman, Yoav and Li, Yiming and Ries, Jonas},
  doi       = {10.1038/s41592-024-02282-x},
  issn      = {1548-7105},
  journal   = {Nat. Methods},
  month     = jun,
  number    = {6},
  pages     = {1082--1093},
  publisher = {Springer Science and Business Media LLC},
  title     = {Universal inverse modeling of point spread functions for SMLM localization and microscope characterization},
  volume    = {21},
  year      = {2024}
}

@article{Liu2025Noise,
  author    = {Liu, Yiming and Panezai, Spozmai and Wang, Yutong and Stallinga, Sjoerd},
  doi       = {10.1038/s41467-025-56241-x},
  issn      = {2041-1723},
  journal   = {Nat. Commun.},
  month     = jan,
  number    = {1},
  publisher = {Springer Science and Business Media LLC},
  title     = {Noise amplification and ill-convergence of Richardson-Lucy deconvolution},
  volume    = {16},
  year      = {2025}
}

@article{Luengo2020survey,
  author    = {Luengo, David and Martino, Luca and Bugallo, Mónica and Elvira, Víctor and Särkkä, Simo},
  day       = {29},
  doi       = {10.1186/s13634-020-00675-6},
  issn      = {1687-6180},
  journal   = {EURASIP J. Adv. Signal Process.},
  month     = may,
  number    = {1},
  pages     = {25},
  publisher = {Springer Science and Business Media LLC},
  title     = {A survey of Monte Carlo methods for parameter estimation},
  volume    = {2020},
  year      = {2020}
}

@article{Manley2008High,
  author    = {Manley, Suliana and Gillette, Jennifer M and Patterson, George H and Shroff, Hari and Hess, Harald F and Betzig, Eric and Lippincott-Schwartz, Jennifer},
  doi       = {10.1038/nmeth.1176},
  issn      = {1548-7105},
  journal   = {Nat. Methods},
  month     = jan,
  number    = {2},
  pages     = {155--157},
  publisher = {Springer Science and Business Media LLC},
  title     = {High-density mapping of single-molecule trajectories with photoactivated localization microscopy},
  volume    = {5},
  year      = {2008}
}

@article{Martens2022,
  author    = {Martens, Koen J. A. and Turkowyd, Bartosz and Endesfelder, Ulrike},
  doi       = {10.3389/fbinf.2021.817254},
  issn      = {2673-7647},
  journal   = {Front. Bioinform.},
  month     = feb,
  publisher = {Frontiers Media SA},
  title     = {Raw Data to Results: A Hands-On Introduction and Overview of Computational Analysis for Single-Molecule Localization Microscopy},
  volume    = {1},
  year      = {2022}
}

@article{Moerner2015Single,
  author    = {Moerner, W. E. and Shechtman, Yoav and Wang, Quan},
  doi       = {10.1039/c5fd00149h},
  issn      = {1364-5498},
  journal   = {Faraday Discuss.},
  pages     = {9--36},
  publisher = {Royal Society of Chemistry (RSC)},
  title     = {Single-molecule spectroscopy and imaging over the decades},
  volume    = {184},
  year      = {2015}
}

\end{document}